\documentclass[aps,prd,superscriptaddress,amsmath,amssymb,amsthm,eqsecnum,nofootinbib, preprintnumbers,10pt]{revtex4-2}
\usepackage{url}
\usepackage[colorlinks=true,
    linkcolor=blue,
    citecolor=blue,
    urlcolor=blue]{hyperref}

  \usepackage{array}
\usepackage{graphicx}
\usepackage{epstopdf}
\usepackage{float}
\usepackage{subcaption}
\usepackage{caption}
\usepackage{booktabs}

\usepackage{color}
\usepackage[T1]{fontenc}
\usepackage[utf8]{inputenc}
\usepackage[toc,page]{appendix}
\usepackage[usenames,dvipsnames]{xcolor}
\usepackage[normalem]{ulem}

\usepackage[edges]{forest}

\usepackage{xcolor}
\usepackage{scalerel}

\usepackage{tikz}
\usetikzlibrary{svg.path}
\definecolor{darkmagenta}{HTML}{8B008B}
\definecolor{orcidlogocol}{HTML}{A6CE39}
\tikzset{
  orcidlogo/.pic={
    \fill[orcidlogocol] svg{M256,128c0,70.7-57.3,128-128,128C57.3,256,0,198.7,0,128C0,57.3,57.3,0,128,0C198.7,0,256,57.3,256,128z};
    \fill[white] svg{M86.3,186.2H70.9V79.1h15.4v48.4V186.2z}
                 svg{M108.9,79.1h41.6c39.6,0,57,28.3,57,53.6c0,27.5-21.5,53.6-56.8,53.6h-41.8V79.1z M124.3,172.4h24.5c34.9,0,42.9-26.5,42.9-39.7c0-21.5-13.7-39.7-43.7-39.7h-23.7V172.4z}
                 svg{M88.7,56.8c0,5.5-4.5,10.1-10.1,10.1c-5.6,0-10.1-4.6-10.1-10.1c0-5.6,4.5-10.1,10.1-10.1C84.2,46.7,88.7,51.3,88.7,56.8z};
  }
}

\newcommand\orcidicon[1]{\href{https://orcid.org/#1}{\mbox{\scalerel*{
\begin{tikzpicture}[yscale=-1, scale=0.05, transform shape]
\pic{orcidlogo};
\end{tikzpicture}
}{|}}}}
\makeatletter
\providecommand*{\diff}%
{\@ifnextchar^{\DIfF}{\DIfF^{}}}
\def\DIfF^#1{%
\mathop{\mathrm{\mathstrut d}}%
\nolimits^{#1}\gobblespace}
\def\gobblespace{%
\futurelet\diffarg\opspace}
\def\opspace{%
\let\DiffSpace\!%
\ifx\diffarg(%
\let\DiffSpace\relax
\else
\ifx\diffarg[%
\let\DiffSpace\relax
\else
\ifx\diffarg\{%
\let\DiffSpace\relax
\fi\fi\fi\DiffSpace}

\usepackage{adjustbox}
\usepackage{bm}

\begin{document}


\title{Swirling spacetimes in higher dimensions: \\ Vacuum backgrounds and black holes}
\author{Jos\'e Barrientos\,$^{\orcidicon{0000-0003-3445-8151}}$}
\email{barrientos@math.cas.cz}
\affiliation{Institute of Mathematics of the Czech Academy of Sciences, {\v Z}itn{\'a} 25, 115 67 Praha 1, Czech Republic}
\affiliation{Vicerrector\'ia de Investigaci\'on y Postgrado, Universidad de La Serena, La Serena 1700000, Chile}

\author{Adolfo Cisterna\,$^{\orcidicon{0000-0002-6949-7882}}$}
\email{adolfo.cisterna@mff.cuni.cz}
\affiliation{Sede Esmeralda, Universidad de Tarapac{\'a}, Avenida Luis Emilio Recabarren 2477, Iquique, Chile}
\affiliation{Institute of Theoretical Physics, Faculty of Mathematics and Physics,
Charles University, V Hole{\v s}ovi{\v c}k{\' a}ch 2, 180 00 Praha 8, Czech Republic}
\author{Carlos Herdeiro\,$^{\orcidicon{0000-0002-9619-2013}}$}
\email{herdeiro@ua.pt}
\affiliation{Departamento de Matem\'atica da Universidade de Aveiro and Centre for Research and Development  in Mathematics and Applications (CIDMA), Campus de Santiago, 3810-193 Aveiro, Portugal}%
\author{Konstantinos Pallikaris\,$^{\orcidicon{0000-0003-2403-892X}}$}
\email{konstantinos.pallikaris@gmail.com}
\affiliation{Laboratory of Theoretical Physics, Institute of Physics, University of Tartu, W. Ostwaldi 1, 50411 Tartu, Estonia}%



\begin{abstract}
We construct higher-dimensional swirling spacetimes in vacuum General
Relativity using the hidden symmetries of the dimensionally reduced
theory. Working directly with the coset-space formulation, we perform
in five dimensions a reduction along the two independent axial Killing
vectors, obtaining an $SL(3,\mathbb{R})/SO(3)$ sigma model and a
genuinely five-dimensional background with two independent swirling
parameters. The same transformation yields doubly swirling
Schwarzschild--Tangherlini and Myers--Perry geometries.
A distinctive feature of the new background is its multidirectional
asymptotic structure. Correlated radial-angular limits reveal
directional curvature singularities that are invisible in the usual
large-radius limit at fixed angle. We further exhibit causal geodesics
that reach such singular asymptotic ends in finite affine parameter.
Finally, we describe the maximally axial extension of this construction
to higher dimensions: five dimensions are maximal for an entirely
axial reduction to three dimensions, while six dimensions admit an
additional nontrivial translational deformation, with further
dimensions contributing only flat spectator directions.
\end{abstract}

\maketitle

\section{Introduction}
\label{sec:intro}

Hidden symmetries of dimensionally reduced General Relativity provide a
powerful framework for generating exact solutions of Einstein's
equations. In four-dimensional vacuum gravity, reduction along a
non-null Killing direction leads to a three-dimensional gravity-coupled
nonlinear sigma model with global $SL(2,\mathbb{R})$ symmetry
\cite{Ehlers:1957zz,Maison:1979kt}. The norm and twist of the Killing
vector can equivalently be encoded in the complex Ernst potential
\cite{Ernst:1967by,Ernst:1967wx}, and the resulting finite symmetry
transformations have long been used as solution-generating techniques
\cite{Kinnersley1973,Kinnersley:1977,KinnersleyChitre:1977,
KinnersleyChitre:1978a,KinnersleyChitre:1978b}. A further reduction to
two dimensions exposes the larger integrable structure associated with
the Geroch group and inverse-scattering methods
\cite{Geroch:1970nt,Geroch:1972,Belinski:1978,Belinski:1979}.

The geometrical effect of these transformations depends crucially on the
causal character of the Killing direction used in the reduction. The
timelike realisation of the Ehlers transformation, for example,
generates the NUT sector from a Schwarzschild seed
\cite{Taub:1950ez,Newman:1963yy}. By contrast, a spacelike reduction
along an axial Killing vector can act nontrivially already on Minkowski
spacetime, since the norm of the axial vector is coordinate dependent.
The resulting geometry is the swirling spacetime, originally introduced
by Harrison \cite{Harrison} and subsequently studied in a variety of
contexts
\cite{Garfinkle:1994wk,Contopoulos:2015wra,Astorino:2022aam,
Barrientos:2024pkt,DiPinto:2025yaa,Cisterna:2023uqf,
Capobianco:2023kse,Capobianco:2024jhe,Moreira:2024sjq,
Capobianco:2025lxn,Barrientos:2024umq,Colleaux:2026qew}. Its
characteristic feature is a non-uniform frame dragging whose orientation
changes across the spacetime, despite the absence of a net dipolar
angular momentum.

The higher-dimensional problem is structurally richer. In five
dimensions, reduction to three dimensions requires two commuting Killing
vectors and yields a sigma model with global $SL(3,\mathbb{R})$
symmetry \cite{Maison:1979kt}. Depending on whether the reduced
directions are timelike or spacelike, the relevant coset is
$SL(3,\mathbb{R})/SO(2,1)$ or $SL(3,\mathbb{R})/SO(3)$,
respectively. The former has been extensively used in the construction
of rotating five-dimensional solutions
\cite{Maison:1979kt,Giusto:2007fx,Suzuki:2024eoz}, whereas the purely
spacelike sector has received comparatively less attention.

A further subtlety arises because the higher-dimensional Ernst-like
formulation is more restrictive than the full coset description. In the
known five-dimensional Einstein--Maxwell formulation, one of the
spacelike Killing directions is required to remain hypersurface
orthogonal, so that only one independent twist is retained
\cite{Yazadjiev:2008pt}. This restriction was inherited by the recent
five-dimensional swirling construction of
Ref.~\cite{Vigano:2026hqd}, where a single swirling sector was
activated.

In the present work we instead work directly with the full coset-space
formulation and reduce five-dimensional vacuum gravity along the two
independent axial Killing vectors associated with the two orthogonal
rotation planes. The resulting $SL(3,\mathbb{R})/SO(3)$ sigma model
admits two independent Ehlers parameters, leading to a genuinely
five-dimensional doubly swirling background. This construction is
inequivalent to the axial-translational background considered in
Ref.~\cite{Vigano:2026hqd}: although both start from five-dimensional
Minkowski spacetime, the corresponding reduced sigma model data differ
because the Killing directions entering the reduction have different
geometrical character. In particular, the single-swirl limit of the
axial-axial construction does not reduce to the axial-translational
background.

The same transformation can be applied to black hole seeds. We construct
the doubly swirling Schwarzschild--Tangherlini geometry and analyse the
deformation of its horizon, as well as its rotational and thermodynamic
properties. We also give the corresponding transformation of the general
doubly rotating Myers--Perry solution, which combines two intrinsic
rotation parameters with two independent background swirls. A detailed
study of the latter geometry will be presented separately
\cite{toappear}.

A central feature of the new background is its asymptotic structure.
The limit $r\to\infty$ is intrinsically multidirectional: different
correlated radial-angular scalings lead to inequivalent asymptotic
behaviour. Although the Kretschmann invariant decays along fixed
interior angles, it diverges along particular asymptotic directions.
These directional singularities are therefore invisible in the
conventional large-radius limit at fixed angle. We further show that
this behaviour is physically relevant by exhibiting null and timelike
geodesics that reach a curvature-singular asymptotic end in finite
affine parameter. The corresponding incompleteness cannot thus be
cured by extending the coordinate patch.

We finally examine the maximally axial extension of the construction to
higher dimensions. Five dimensions are the highest dimension in which
the reduction to three dimensions can be performed entirely along axial
Killing directions. In six and higher dimensions, a purely spacelike
reduction necessarily involves translational directions. Their Ehlers
parameters produce a genuine translational deformation, but rotational
symmetry among the flat directions reduces them to a single effective
parameter; the remaining dimensions are flat spectators. Attempting to
activate a third independent rotational plane instead requires a
reduction containing the timelike Killing vector and, for the Ehlers
transformation considered here, restricts the admissible radial domain.
Thus the five-dimensional solution represents the maximally axial,
globally defined member of this class.

The paper is organised as follows. In \autoref{sec2} we review the
dimensional-reduction formalism and the relevant Ehlers transformations.
In \autoref{sec3} we construct the five-dimensional doubly swirling
background and analyse its rotational and asymptotic properties. In
\autoref{sec4} we present the doubly swirling
Schwarzschild--Tangherlini geometry and discuss its horizon properties
and thermodynamics. The higher-dimensional extension is developed in
\autoref{sec5}, and our conclusions are collected in
\autoref{sec6}. The general doubly swirling Myers--Perry solution is
given in Appendix~\ref{app:swirlingMP}.

\section{Preliminaries}\label{sec2}

We briefly review the dimensional-reduction framework used to generate
the solutions presented below. Working directly with the nonlinear
sigma model variables makes the hidden symmetry group manifest and, in
particular, allows us to treat different choices of commuting Killing
directions within a common formalism.

Consider $D$-dimensional vacuum General Relativity with metric
$\bm{g}=g_{MN}\diff x^M\otimes\diff x^N$ and action
\begin{equation}
    S=\frac{1}{2\kappa_D^2}
    \int \diff^D\!x\,\sqrt{-g}\,R .
    \label{eq:Ddim-EH-action}
\end{equation}
We assume the existence of $D-3$ mutually commuting Killing vector
fields
\begin{equation}
    \bm{\xi}_a=\partial_{x^a},
    \qquad
    a=1,\ldots,D-3,
\end{equation}
and denote by
\begin{equation}
    f_{ab}=g(\bm{\xi}_a,\bm{\xi}_b),
    \qquad
    f=\det(f_{ab}),
\end{equation}
their Gram matrix and its determinant. In coordinates adapted to these
isometries, the metric can be written as
\begin{equation}
    \mathrm{ds}^2
    =
    f_{ab}
    \bigl(\diff x^a+\mathcal{A}^a\bigr)
    \bigl(\diff x^b+\mathcal{A}^b\bigr)
    +\frac{1}{|f|}\,
    h_{ij}\diff y^i\diff y^j ,
    \label{eq:KKansatz}
\end{equation}
where $y^i$, $i=1,2,3$, are coordinates on the orbit space,
$\bm{h}=h_{ij}\diff y^i\otimes\diff y^j$ is its metric, and
\begin{equation}
    \mathcal{A}^a=\mathcal{A}^a{}_i\diff y^i
\end{equation}
are the Kaluza--Klein gauge 1-forms. All reduced fields depend only on
the orbit-space coordinates $y^i$.

Introducing the field strengths
\begin{equation}
    \mathcal{F}^a=\diff\mathcal{A}^a ,
\end{equation}
the mixed components of the vacuum Einstein equations imply
\begin{equation}
    \diff\!\left(
    |f|f_{ab}\star\mathcal{F}^b
    \right)=0 ,
    \label{eq:Iicomponent}
\end{equation}
where $\star$ denotes the Hodge dual with respect to $\bm{h}$. Hence,
locally, one may introduce $D-3$ twist potentials $\chi_a$ satisfying
\begin{equation}
    \diff\chi_a
    =
    |f|f_{ab}\star\mathcal{F}^b .
    \label{eq:PotentialEq}
\end{equation}
Equivalently,
\begin{equation}
    \mathcal{F}^a
    =
    -\frac{1}{f}\,
    f^{ab}\star\diff\chi_b .
    \label{eq:Wstrengths}
\end{equation}

The remaining reduced scalar equations can be encoded in the Maison
matrix
\begin{equation}
    \bm{M}
    =
    \begin{pmatrix}
        \bm{f}+f^{-1}\bm{\chi}\bm{\chi}^{T}
        &
        |f|^{-1}\bm{\chi}
        \\[2mm]
        |f|^{-1}\bm{\chi}^{T}
        &
        f^{-1}
    \end{pmatrix},
\end{equation}
where $\bm{\chi}$ is the column vector with components $\chi_a$.
Defining the matrix-valued 1-form
\begin{equation}
    \bm{J}
    =
    \bm{M}^{-1}\diff\bm{M}
    =
    \bm{J}_i\diff y^i ,
\end{equation}
the reduced Einstein equations take the compact form
\begin{equation}
    \diff\star\bm{J}=0,
    \qquad
    \mathcal{R}_{ij}
    =
    \frac{1}{4}
    \operatorname{Tr}
    \bigl(\bm{J}_i\bm{J}_j\bigr),
    \label{eq:RedFieldEqs}
\end{equation}
where $\mathcal{R}_{ij}$ is the Ricci tensor of the orbit-space metric.
These equations follow from the three-dimensional action
\begin{equation}
    S_3
    =
    \int \diff^3y\,\sqrt{|h|}
    \left[
        \mathcal{R}
        -\frac{1}{8}
        \operatorname{Tr}
        \bigl(\bm{J}_i\bm{J}_j\bigr)h^{ij}
    \right].
    \label{eq:RedAction}
\end{equation}

The reduced theory is therefore three-dimensional gravity coupled to a
nonlinear sigma model with target space
\begin{equation}
    \frac{\mathrm{SL}(D-2,\mathbb{R})}{H},
    \qquad
    H=
    \begin{cases}
        \mathrm{SO}(D-4,2), & f<0,\\[1mm]
        \mathrm{SO}(D-2),   & f>0.
    \end{cases}
\end{equation}
The sign of $f$ distinguishes reductions containing one timelike
Killing direction from purely spacelike reductions. The global
$\mathrm{SL}(D-2,\mathbb{R})$ symmetry acts on the Maison matrix by
congruence transformations and leaves both
\eqref{eq:RedAction} and \eqref{eq:RedFieldEqs} invariant.

For later use, it is convenient to recall three elementary classes of
finite transformations. Constant shifts of the twist potentials,
\begin{equation}
    \bm{f}'=\bm{f},
    \qquad
    \bm{\chi}'=\bm{\chi}+\bm{c},
    \label{eq:ShiftFreedom}
\end{equation}
with constant $\bm{c}$, do not change the spacetime metric. A second
class is generated by constant invertible matrices $\bm{C}$,
\begin{equation}
    \bm{f}'=\bm{C}^{T}\bm{f}\bm{C},
    \qquad
    \bm{\chi}'=
    C\,\bm{C}^{T}\bm{\chi},
    \qquad
    C\equiv\det\bm{C}.
\end{equation}
Up to the redefinition
\begin{equation}
    x'^a=C\,C^a{}_b x^b ,
\end{equation}
these transformations amount to a homothety of the spacetime metric.
For $C=1$, they reduce to a linear change of basis in the Abelian
Killing algebra.

The transformations relevant to the present work are the Ehlers
transformations. They are parametrised by a constant real vector
\begin{equation}
    j^a=(j_1,\ldots,j_{D-3})^a
\end{equation}
and act according to
\begin{equation}
    f'_{ab}
    =
    f_{ab}
    +\frac{1}{f}\chi_a\chi_b
    -\Pi\,\chi'_a\chi'_b ,
\end{equation}
with
\begin{equation}
    \chi'_a
    =
    \frac{1}{\Pi}
    \left[
        f_{ab}j^b
        +\frac{1}{f}\chi_a
        \bigl(1+j^b\chi_b\bigr)
    \right],
\end{equation}
and
\begin{equation}
    \Pi
    =
    f_{ab}j^a j^b
    +\frac{1}{f}
    \bigl(1+j^a\chi_a\bigr)^2 .
\end{equation}
Since
\begin{equation}
    f'=\Pi^{-1},
\end{equation}
preservation of the signature of the reduced Killing sector requires
\begin{equation}
    f\Pi>0 .
    \label{eq:EhlersPositivity}
\end{equation}
The Ehlers transformations form a $(D-3)$-dimensional Abelian subgroup
of $\mathrm{SL}(D-2,\mathbb{R})$. The condition
\eqref{eq:EhlersPositivity} will become particularly relevant when
reductions containing a timelike Killing direction are considered in
\autoref{sec5}.

The solution-generating procedure is then straightforward. Starting
from a seed metric in the form \eqref{eq:KKansatz}, one determines
$\bm{f}$, $\bm{h}$ and $\mathcal{A}^a$, integrates
\eqref{eq:PotentialEq} to obtain the twist potentials, and applies an
Ehlers transformation to construct $\bm{f}'$ and $\bm{\chi}'$. The
orbit-space metric $\bm{h}$ is unchanged by the transformation, while
the transformed Kaluza--Klein fields $\mathcal{A}'^a$ are recovered
from \eqref{eq:Wstrengths}. The full transformed spacetime then follows
from the Kaluza--Klein ansatz \eqref{eq:KKansatz}.

\section{The five-dimensional swirling spacetime}
\label{sec3}

\subsection{The background geometry}
\label{subsec:background-geometry}

We begin with the five-dimensional member of the maximally axial
swirling family. As discussed in \autoref{sec5}, five dimensions are
special in that the reduction to three dimensions can be performed
entirely along the two independent axial Killing directions. In the
absence of translational Ehlers deformations, its higher-dimensional
embedding is simply
\begin{equation}
    \mathcal{M}^{\text{2-s}}_5\times\mathbb{R}^{D-5},
\end{equation}
where $\mathcal{M}^{\text{2-s}}_5$ denotes the doubly swirling
five-dimensional geometry constructed below.

Using spherical coordinates adapted to the two orthogonal rotation
planes, the metric takes the form
\begin{equation}
\label{eq:SwMet}
\begin{split}
\mathrm{ds}^2={}&
\frac{1+(j_2 r^3\cos^2\theta\sin\theta)^2}{\Lambda^2}
r^2\sin^2\theta\,\bm{\Theta}_1^2
+
\frac{1+(j_1 r^3\cos\theta\sin^2\theta)^2}{\Lambda^2}
r^2\cos^2\theta\,\bm{\Theta}_2^2
\\
&-
\frac{2j_1j_2(r^2\cos\theta\sin\theta)^4}{\Lambda^2}
\bm{\Theta}_1\bm{\Theta}_2
+\Lambda^2
\left(
-\diff t^2+\diff r^2+r^2\diff\theta^2
\right),
\end{split}
\end{equation}
where
\begin{equation}
\begin{split}
\bm{\Theta}_1 &=
\diff\phi_1
+j_1r^2
\left(
2\cos^2\theta-\sin^2\theta
\right)\diff t,
\\
\bm{\Theta}_2 &=
\diff\phi_2
+j_2r^2
\left(
\cos^2\theta-2\sin^2\theta
\right)\diff t,
\end{split}
\label{eq:SwTheta}
\end{equation}
and
\begin{equation}
\Lambda^2
=
1+r^6\sin^2\theta\cos^2\theta
\left(
j_1^2\sin^2\theta+j_2^2\cos^2\theta
\right).
\label{eq:SwLambda}
\end{equation}
The two swirl parameters introduce the characteristic length scales
\begin{equation}
    \ell_a=|j_a|^{-1/3},
    \qquad a=1,2.
\end{equation}

Two special sectors of \eqref{eq:SwMet} are worth displaying
explicitly. First, consider the single-swirl limit $j_2=0$. The metric
reduces to
\begin{equation}
\label{eq:single-swirl}
\mathrm{ds}^2={}
\frac{r^2\sin^2\theta}
{1+j_1^2r^6\cos^2\theta\sin^4\theta}
\,\bm{\Theta}_1^2
+r^2\cos^2\theta\,\diff\phi_2^2
+
\left(
1+j_1^2r^6\cos^2\theta\sin^4\theta
\right)
\left(
-\diff t^2+\diff r^2+r^2\diff\theta^2
\right),
\end{equation}
with
\begin{equation}
    \bm{\Theta}_1
    =
    \diff\phi_1
    +j_1r^2
    \left(
    2\cos^2\theta-\sin^2\theta
    \right)\diff t .
\end{equation}
Although only one Ehlers parameter is nonzero, this geometry remains
intrinsically five-dimensional. In particular, the second axial sector
does not become a flat spectator direction: its radius
$r\cos\theta$ continues to enter the deformation factor multiplying
the three-dimensional orbit-space metric. Consequently,
\eqref{eq:single-swirl} is inequivalent to the direct product of the
four-dimensional swirling background with a flat line. This is the
geometrical distinction between the present axial-axial reduction and
the axial-translational construction of
Ref.~\cite{Vigano:2026hqd}.

A second distinguished sector is obtained when the two swirls have
equal strength,
\begin{equation}
    j_1=j_2\equiv j .
\end{equation}
There is then a single characteristic scale
$\ell=|j|^{-1/3}$, and the deformation function simplifies to
\begin{equation}
    \Lambda_{\rm eq}^2
    =
    1+j^2r^6\sin^2\theta\cos^2\theta .
    \label{eq:equal-swirl-Lambda}
\end{equation}
It is useful in this case to write the metric as
\begin{equation}
\label{eq:equal-swirl}
\begin{split}
\mathrm{ds}^2={}&
\Lambda_{\rm eq}^2
\left(
-\diff t^2+\diff r^2+r^2\diff\theta^2
\right)
+r^2\sin^2\theta\,\bm{\Theta}_1^2
+r^2\cos^2\theta\,\bm{\Theta}_2^2
\\
&-
\frac{j^2r^8\sin^2\theta\cos^2\theta}
{\Lambda_{\rm eq}^2}
\left(
\sin^2\theta\,\bm{\Theta}_1
+\cos^2\theta\,\bm{\Theta}_2
\right)^2 ,
\end{split}
\end{equation}
where
\begin{equation}
\begin{split}
\bm{\Theta}_1 &=
\diff\phi_1
+jr^2(3\cos^2\theta-1)\diff t,
\\
\bm{\Theta}_2 &=
\diff\phi_2
+jr^2(1-3\sin^2\theta)\diff t .
\end{split}
\label{eq:equal-swirl-Theta}
\end{equation}
The equality of the two swirl parameters makes the two rotational
planes equivalent and enhances the discrete symmetry of the geometry.
In particular, \eqref{eq:equal-swirl} is invariant under
\begin{equation}
    \mathcal{I}:
    \qquad
    (t,\theta,\phi_1,\phi_2)
    \longmapsto
    \left(
    -t,\frac{\pi}{2}-\theta,\phi_2,\phi_1
    \right),
    \label{eq:equal-swirl-symmetry}
\end{equation}
under which
$\bm{\Theta}_1\leftrightarrow\bm{\Theta}_2$.
Thus, in the equal-swirl sector the interchange of the two orthogonal
rotation planes becomes an actual $\mathbb{Z}_2$ isometry of a fixed
solution, rather than merely a map between solutions with interchanged
parameters. Notice, however, that the dependence of
$\Lambda_{\rm eq}$ on $\theta$ remains nontrivial. The enhancement is
therefore discrete and does not render the spacetime cohomogeneity one.

For generic nonzero $(j_1,j_2)$, the continuous isometry group is
generated by the commuting Killing vector fields
$\partial_t$, $\partial_{\phi_1}$, and $\partial_{\phi_2}$.
The two symmetry axes are located at $\theta=0$ and
$\theta=\pi/2$. Introducing Cartesian coordinates on each rotational
plane,
\begin{equation}
\label{eq:CartesianMark}
\partial_{\phi_a}
=
\hat{x}_a\partial_{\hat{y}_a}
-\hat{y}_a\partial_{\hat{x}_a},
\end{equation}
shows that the corresponding axial Killing field vanishes at
$\hat{x}_a=\hat{y}_a=0$.

The axes are regular. Their local geometries are
\begin{equation}
{\mathrm{ds}}^2\sim
\begin{cases}
-\diff t^2+\diff r^2
+\diff\hat{x}_1^2+\diff\hat{y}_1^2
+r^2\bigl[\diff\phi_2+j_2r^2\diff t\bigr]^2,
& \theta\to0,
\\[1mm]
-\diff t^2+\diff r^2
+\diff\hat{x}_2^2+\diff\hat{y}_2^2
+r^2\bigl[\diff\phi_1-j_1r^2\diff t\bigr]^2,
& \theta\to\pi/2,
\end{cases}
\end{equation}
with the standard $2\pi$ periodicities for $\phi_1$ and $\phi_2$.
The origin is also regular,
\begin{equation}
\mathrm{ds}^2
\underset{r\to0}{\sim}
-\diff t^2
+\diff\hat{x}_1^2+\diff\hat{y}_1^2
+\diff\hat{x}_2^2+\diff\hat{y}_2^2 .
\end{equation}

Finally, the coordinate $t$ is a global temporal function. Indeed,
\begin{equation}
    g^{tt}=-\Lambda^{-2}<0 ,
\end{equation}
so its gradient is everywhere timelike. The spacetime is therefore
stably causal~\cite{Hawking1969}, and in particular contains no closed
causal curves.

\subsection{Rotational structure and conserved charges}
\label{subsec:background-rotation}

The swirling background exhibits nontrivial frame dragging in both
rotational planes. Since no stationary Killing vector is timelike in
all asymptotic directions, there is no globally preferred stationary
frame. It is therefore useful to consider the family
\begin{equation}
    \bm{k}
    =
    \partial_t+\omega^a\partial_{\phi_a},
    \qquad
    \omega^a=(\omega_1,\omega_2)^a,
\end{equation}
with constant $\omega^a$.

Consider a stationary observer with five-velocity
\begin{equation}
    \bm{u}
    \propto
    \bm{k}
    +
    \Omega^a\partial_{\phi_a}.
\end{equation}
For a zero-angular-momentum observer (ZAMO), the conditions
$u_{\phi_a}=0$ give
\begin{equation}
    k^M(\partial_{\phi_a})^N g_{MN}
    +
    f_{ab}\Omega^b
    =
    0,
\end{equation}
and hence
\begin{equation}
    \Omega^a
    =
    -f^{ab}
    k^M(\partial_{\phi_b})^N g_{MN}=(\Omega_1,\Omega_2)^a.
\end{equation}
For the metric \eqref{eq:SwMet},
\begin{equation}
\begin{split}
    \Omega_1
    &=
    -j_1r^2
    \left(
        2\cos^2\theta-\sin^2\theta
    \right)
    -\omega_1,
    \\
    \Omega_2
    &=
    -j_2r^2
    \left(
        \cos^2\theta-2\sin^2\theta
    \right)
    -\omega_2.
\end{split}
\label{eq:ZAMO-angular-velocities}
\end{equation}
Changing the stationary Killing field therefore shifts each angular
velocity by a constant. The spatial variation of the dragging is
independent of this choice. In particular, defining
\begin{equation}
    \Delta\Omega^a
    =
    \Omega^a-\Omega^a_0,
\end{equation}
with respect to any fixed reference point, removes this frame
ambiguity.

Choosing the regular origin as reference and introducing double-polar
coordinates
\begin{equation}
    r_1=r\sin\theta,
    \qquad
    r_2=r\cos\theta,
\end{equation}
one obtains
\begin{equation}
    \Delta\Omega_1
    =
    j_1\left(r_1^2-2r_2^2\right),
    \qquad
    \Delta\Omega_2
    =
    j_2\left(2r_1^2-r_2^2\right).
    \label{eq:dragging-double-polar}
\end{equation}
For nonzero $j_1$ and $j_2$, the two rotational sectors reverse their
sense of dragging across the hypersurfaces
\begin{equation}
    r_1=\sqrt{2}\,r_2,
    \qquad
    r_2=\sqrt{2}\,r_1,
    \label{eq:dragging-reversal}
\end{equation}
respectively. Equivalently,
\begin{equation}
    \tan^2\theta=2,
    \qquad
    \tan^2\theta=\frac{1}{2}.
\end{equation}
These surfaces are independent of the magnitude of the swirl
parameters and are symmetrically placed about $\theta=\pi/4$.
For $j_1,j_2>0$, the sign structure is
\begin{equation}
\begin{array}{c|cc}
\hline
\text{Region} & \Delta\Omega_1 & \Delta\Omega_2 \\
\hline
r_1 < r_2/\sqrt{2}
& - & - \\[1mm]
r_2/\sqrt{2} < r_1 < \sqrt{2}\,r_2
& - & + \\[1mm]
r_1 > \sqrt{2}\,r_2
& + & + \\
\hline
\end{array}.
\label{eq:dragging-signs}
\end{equation}
Thus the intermediate region is characterised by opposite senses of
dragging in the two rotational planes. Reversing the sign of either
$j_a$ reverses the corresponding column in
\eqref{eq:dragging-signs}. In the equal-swirl sector discussed in
\autoref{subsec:background-geometry}, the two reversal surfaces are
mapped into one another by the enhanced plane-exchange symmetry.

Away from the sign-reversal hypersurfaces, the dragging grows
quadratically at large radius,
\begin{equation}
    |\Delta\Omega^a|
    =
    \mathcal{O}(|j^a|r^2),
    \qquad
    r\rightarrow\infty.
\end{equation}
This behaviour persists throughout the generic correlated asymptotic
limits discussed in \autoref{subsec:background-asymptotics}, and shows that the
background possesses an unbounded two-component frame-dragging
structure.

\begin{figure}[t]
    \centering

    \begin{subfigure}[t]{0.45\textwidth}
        \centering
        \includegraphics{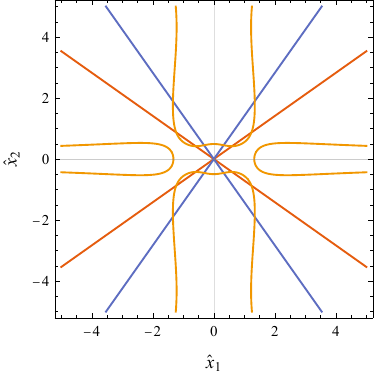}
        \caption{$\bar{j}_1=1$ and $\bar{j}_2=0.1$}
    \end{subfigure}
    \quad
    \begin{subfigure}[t]{0.45\textwidth}
        \centering
        \includegraphics{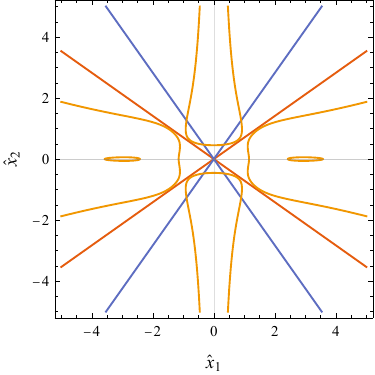}
        \caption{$\bar{j}_1=0.1$ and $\bar{j}_2=1$}
    \end{subfigure}

    \vspace{1em}

    \begin{subfigure}[t]{0.45\textwidth}
        \centering
        \includegraphics{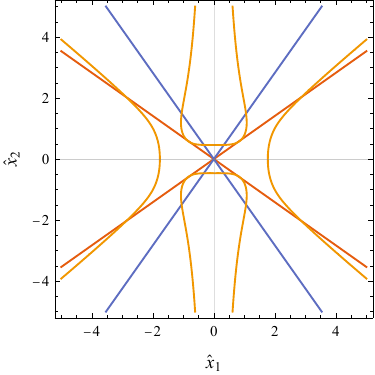}
        \caption{$\bar{j}_1=0.5$ and $\bar{j}_2=0.75$}
    \end{subfigure}
    \quad
    \begin{subfigure}[t]{0.45\textwidth}
        \centering
        \includegraphics{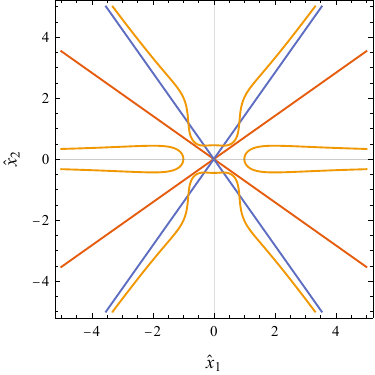}
        \caption{$\bar{j}_1=2$ and $\bar{j}_2=1$}
    \end{subfigure}

    \caption{
    Cross-sections at $\hat{y}_1=\hat{y}_2=0$ of the ergosurface
    (yellow) and of the sign-reversal surfaces
    $\Delta\Omega_1=0$ (red) and $\Delta\Omega_2=0$ (blue).
    The ergo-geometry is defined with respect to
    $\bm{k}=\partial_t+\partial_{\phi_1}
    +2\partial_{\phi_2}$, while angular-velocity differences are
    measured relative to the origin. The coordinates and swirl
    parameters are displayed in the dimensionless variables defined
    above.
    }
    \label{fig:Ergo-geometry}
\end{figure}

The same absence of a preferred stationary frame affects the notion of
an ergoregion. With respect to a stationary Killing vector $\bm{k}$, the latter is defined as the collection of spacetime points $p$ at which $\bm{k}$ is spacelike, namely
\begin{equation}
    \mathcal{E}_{\bm{k}}
    =
    \left\{
        p\;|\;g(\bm{k},\bm{k})>0
    \right\}.
\end{equation}
Using the Kaluza--Klein decomposition,
\begin{equation}
    g(\bm{k},\bm{k})
    =
    -\Lambda^2
    +
    f_{ab}
    \left(
        \mathcal{A}^a+\omega^a
    \right)
    \left(
        \mathcal{A}^b+\omega^b
    \right).
    \label{eq:k-norm}
\end{equation}
For any fixed $\omega^a$, the terms depending explicitly on
$\omega^a$ are subleading in the asymptotic regions analysed below, so
that
\begin{equation}
    g(\bm{k},\bm{k})
    \sim
    g_{tt}
    =
    -\Lambda^2
    +
    f_{ab}\mathcal{A}^a\mathcal{A}^b .
\end{equation}
There exist asymptotic directions for which $g_{tt}>0$ whenever at
least one swirl parameter is nonzero. Hence, although the finite-radius
ergosurfaces depend on the chosen stationary frame, every such fixed
frame contains directions in which its stationary Killing vector
becomes spacelike arbitrarily far from the origin. This behaviour is
illustrated in \autoref{fig:Ergo-geometry}.

For the plots, we introduce an arbitrary reference length $L$ and use
the dimensionless variables
\begin{equation}
    \bar{x}_a=\frac{\hat{x}_a}{L},
    \qquad
    \bar{j}_a=j_aL^3.
\end{equation}

Despite this rotational behaviour, the background carries no Komar
charges. Indeed, for any Killing field $\bm{\xi}$ of a vacuum
spacetime,
\begin{equation}
    \diff\ast\diff\bm{\xi}^{\flat}=0.
\end{equation}
Therefore the Komar integral
\begin{equation}
    Q_{\bm{\xi}}[\mathcal{S}]
    \propto
    \int_{\mathcal{S}}
    \ast\diff\bm{\xi}^{\flat}
\end{equation}
is unchanged under continuous deformations of a closed
three-surface $\mathcal{S}$ within a given homology class. Since the
constant-$t$ slices of the present background are topologically
$\mathbb{R}^4$, every such surface can be contracted to the regular
origin. There the geometry approaches five-dimensional Minkowski
spacetime, and hence
\begin{equation}
    Q_{\bm{\xi}}=0 .
\end{equation}
In particular, both Komar angular momenta vanish. The frame dragging is
therefore an intrinsic property of the background rather than the
field of a source carrying net angular momentum.

\subsection{Asymptotic structure and curvature singularities}
\label{subsec:background-asymptotics}

The asymptotic region of the five-dimensional swirling background is
intrinsically multidirectional. Taking $r\to\infty$ does not determine
a unique leading geometry unless the simultaneous behaviour of
$\theta$ is specified. At a fixed interior angle,
$\theta\to\theta_0\in(0,\pi/2)$, both rotation-plane radii
\begin{equation}
    r_1=r\sin\theta,
    \qquad
    r_2=r\cos\theta,
\end{equation}
diverge at comparable rates. By contrast, correlated limits towards
$\theta=0$ or $\theta=\pi/2$ allow one of the two radii to become
parametrically smaller than the other and lead to different
asymptotic balances.

This feature is specific to the coupled axial-axial geometry. In
particular, a large-radius limit taken with the remaining coordinates
fixed, as in Ref.~\cite{Vigano:2026hqd}, does not probe such correlated
approaches to the angular endpoints and therefore does not by itself
establish bounded curvature throughout the asymptotic domain.

The relevant hierarchy follows directly from the deformation function,
\begin{equation}
\label{eq:asymptotic-lambda}
    \Lambda^2
    =
    1
    +j_1^2r^6\sin^4\theta\cos^2\theta
    +j_2^2r^6\sin^2\theta\cos^4\theta .
\end{equation}
For a fixed interior angle we obtain region (i). Near $\theta=0$, the
two independent scales are
\begin{equation}
    X=r^3\sin^2\theta,
    \qquad
    Y=r^3\sin\theta,
    \qquad
    X=Y\sin\theta .
\end{equation}
The limits $X\to\infty$, $X\to C>0$, and $X\to0$ define regions
(ii-a), (ii-b), and (ii-c), respectively. In the last case, $Y$ may
diverge, approach a finite nonzero constant, or vanish, giving the
three subregions (ii-c-1), (ii-c-2), and (ii-c-3). The corresponding
classification near $\theta=\pi/2$ is obtained by interchanging
$\sin\theta$ and $\cos\theta$. The eleven resulting scaling regimes
are summarised in \autoref{tab:asymptotic-regions}. They represent
different asymptotic balances, not distinct asymptotic ends.

\begin{table}[t]
\centering
\small
\renewcommand{\arraystretch}{1.35}
\setlength{\tabcolsep}{7pt}

\begin{tabular}{cccc}
\toprule
Region
& Angular limit
& First condition
& Additional condition \\
\midrule

(i)
& $\theta\to\theta_0\in(0,\pi/2)$
& ---
& --- \\

\midrule

(ii-a)
& $\theta\to0$
& $r^3\sin^2\theta\to\infty$
& --- \\

(ii-b)
& $\theta\to0$
& $r^3\sin^2\theta\to C>0$
& --- \\

(ii-c-1)
& $\theta\to0$
& $r^3\sin^2\theta\to0$
& $r^3\sin\theta\to\infty$ \\

(ii-c-2)
& $\theta\to0$
& $r^3\sin^2\theta\to0$
& $r^3\sin\theta\to C>0$ \\

(ii-c-3)
& $\theta\to0$
& $r^3\sin^2\theta\to0$
& $r^3\sin\theta\to0$ \\

\midrule

(iii-a)
& $\theta\to\pi/2$
& $r^3\cos^2\theta\to\infty$
& --- \\

(iii-b)
& $\theta\to\pi/2$
& $r^3\cos^2\theta\to C>0$
& --- \\

(iii-c-1)
& $\theta\to\pi/2$
& $r^3\cos^2\theta\to0$
& $r^3\cos\theta\to\infty$ \\

(iii-c-2)
& $\theta\to\pi/2$
& $r^3\cos^2\theta\to0$
& $r^3\cos\theta\to C>0$ \\

(iii-c-3)
& $\theta\to\pi/2$
& $r^3\cos^2\theta\to0$
& $r^3\cos\theta\to0$ \\

\bottomrule
\end{tabular}

\caption{
Classification of the asymptotic scaling regimes of the
five-dimensional doubly swirling background. In every case
$r\to\infty$. The constant $C>0$ is finite and may differ between
branches. For the nested subregions, the conditions in the last two
columns hold simultaneously.
}
\label{tab:asymptotic-regions}
\end{table}

To test the regularity of these limits, we evaluate the exact
Kretschmann invariant
\begin{equation}
    \mathcal{K}
    \equiv
    R_{\mu\nu\rho\sigma}R^{\mu\nu\rho\sigma}.
\end{equation}
Its explicit expression is lengthy and is given in Appendix
\ref{app:kretschmann}. The common denominator is
$\Lambda^{12}$ and therefore never vanishes at finite radius. This,
however, does not guarantee bounded curvature at infinity, where the
result depends on the relative radial and angular scalings.

For compactness, define
\begin{equation}
\label{eq:asymptotic-curvature-functions}
    F(x)
    =
    \frac{(3x-1)(x-3)}{(1+x)^6},
    \qquad
    \mathcal{A}(\theta_0)
    =
    \lim_{r\to\infty}
    r^{16}\mathcal{K}(r,\theta_0).
\end{equation}
For fixed nonzero $j_1,j_2$ and
$\theta_0\in(0,\pi/2)$, the coefficient
$\mathcal{A}(\theta_0)$ is finite and strictly positive. The
asymptotic behaviour of $\mathcal{K}$ in the eleven regions is
summarised in \autoref{tab:kretschmann-eleven-regions}.

\begin{table}[t]
\centering
\small
\setlength{\tabcolsep}{5pt}
\renewcommand{\arraystretch}{1.3}

\begin{tabular}{c c c l}
\toprule
\textbf{Region}
&
\shortstack{
\textbf{Angular limit and}\\
\textbf{correlated conditions}
}
&
\shortstack{
\textbf{Leading behaviour}\\
\textbf{of $\mathcal K$}
}
&
\textbf{Curvature limit}
\\
\midrule

\textbf{(i)}
&
$\theta\to\theta_0\in(0,\pi/2)$
&
$\displaystyle
\mathcal K\sim\frac{\mathcal A(\theta_0)}{r^{16}}
$
&
$0$
\\[3mm]

\textbf{(ii-a)}
&
$\begin{gathered}
\theta\to0,\\
r^3\sin^2\theta\to\infty
\end{gathered}$
&
$\displaystyle
\mathcal K\sim
\frac{72}{j_2^4r^{16}\sin^8\theta}
$
&
$0$
\\[4mm]

\textbf{(ii-b)}
&
$\begin{gathered}
\theta\to0,\\
r^3\sin^2\theta\to C
\end{gathered}$
&
$\displaystyle
\mathcal K\sim
\frac{72}{j_2^4C^4r^4}
$
&
$0$
\\[4mm]

\textbf{(ii-c-1)}
&
$\begin{gathered}
\theta\to0,\\
r^3\sin^2\theta\to0,\\
r^3\sin\theta\to\infty
\end{gathered}$
&
$\displaystyle
\mathcal K\sim
\frac{72}{j_2^4(r^2\sin\theta)^8}
$
&
\shortstack[l]{
Depends on\\
$r^2\sin\theta$
}
\\[4mm]

\textbf{(ii-c-2)}
&
$\begin{gathered}
\theta\to0,\\
r^3\sin^2\theta\to0,\\
r^3\sin\theta\to C
\end{gathered}$
&
$\displaystyle
\frac{\mathcal K}{r^8}
\to
24j_2^4F(j_2^2C^2)
$
&
\shortstack[l]{
Generically divergent;\\
exceptional zeros of $F$\\
require refinement
}
\\[4mm]

\textbf{(ii-c-3)}
&
$\begin{gathered}
\theta\to0,\\
r^3\sin^2\theta\to0,\\
r^3\sin\theta\to0
\end{gathered}$
&
$\displaystyle
\mathcal K\sim72j_2^4r^8
$
&
$+\infty$
\\[3mm]

\midrule

\textbf{(iii-a)}
&
$\begin{gathered}
\theta\to\pi/2,\\
r^3\cos^2\theta\to\infty
\end{gathered}$
&
$\displaystyle
\mathcal K\sim
\frac{72}{j_1^4r^{16}\cos^8\theta}
$
&
$0$
\\[4mm]

\textbf{(iii-b)}
&
$\begin{gathered}
\theta\to\pi/2,\\
r^3\cos^2\theta\to C
\end{gathered}$
&
$\displaystyle
\mathcal K\sim
\frac{72}{j_1^4C^4r^4}
$
&
$0$
\\[4mm]

\textbf{(iii-c-1)}
&
$\begin{gathered}
\theta\to\pi/2,\\
r^3\cos^2\theta\to0,\\
r^3\cos\theta\to\infty
\end{gathered}$
&
$\displaystyle
\mathcal K\sim
\frac{72}{j_1^4(r^2\cos\theta)^8}
$
&
\shortstack[l]{
Depends on\\
$r^2\cos\theta$
}
\\[4mm]

\textbf{(iii-c-2)}
&
$\begin{gathered}
\theta\to\pi/2,\\
r^3\cos^2\theta\to0,\\
r^3\cos\theta\to C
\end{gathered}$
&
$\displaystyle
\frac{\mathcal K}{r^8}
\to
24j_1^4F(j_1^2C^2)
$
&
\shortstack[l]{
Generically divergent;\\
exceptional zeros of $F$\\
require refinement
}
\\[4mm]

\textbf{(iii-c-3)}
&
$\begin{gathered}
\theta\to\pi/2,\\
r^3\cos^2\theta\to0,\\
r^3\cos\theta\to0
\end{gathered}$
&
$\displaystyle
\mathcal K\sim72j_1^4r^8
$
&
$+\infty$
\\

\bottomrule
\end{tabular}

\caption{
Asymptotic behaviour of the Kretschmann invariant for fixed nonzero
$j_1$ and $j_2$. In every region $r\to\infty$, and $C>0$ is finite.
The functions $F$ and $\mathcal{A}$ are defined in
\eqref{eq:asymptotic-curvature-functions}.
}
\label{tab:kretschmann-eleven-regions}
\end{table}

Two points in the table require some care. In regions (ii-c-1) and
(iii-c-1), the curvature limit depends on the additional combinations
$r^2\sin\theta$ and $r^2\cos\theta$, respectively. These may diverge,
approach a finite nonzero value, or vanish, and $\mathcal K$
accordingly tends to zero, a finite value, or diverges. In regions
(ii-c-2) and (iii-c-2), the generic $r^8$ divergence is controlled by
$F$. Its zeros at $x=1/3$ and $x=3$ cancel the displayed leading term,
so the corresponding exceptional limits must be resolved at
subleading order. Since $\mathcal K$ is a Lorentzian contraction, it
need not be positive, and by curvature divergence we mean
$|\mathcal K|\to\infty$.

The singular behaviour itself is unambiguous. For fixed nonzero
$j_1,j_2$,
\begin{equation}
    \mathcal K
    \sim
    72j_2^4r^8
    \qquad
    \text{in region (ii-c-3)},
\end{equation}
whereas
\begin{equation}
    \mathcal K
    \sim
    72j_1^4r^8
    \qquad
    \text{in region (iii-c-3)}.
\end{equation}
Thus curvature may decay in the conventional large-radius limit at
fixed interior angle while diverging along correlated approaches to
the asymptotic corners.

It remains to determine whether these singular directions can be
reached by causal geodesics in finite affine parameter. Write the
metric as
\begin{equation}
    \mathrm{ds}^2
    =
    \mathcal{G}_{ij}\diff z^i\diff z^j
    +
    \Lambda^2
    \left(
        \diff r^2+r^2\diff\theta^2
    \right),
    \qquad
    z^i=(t,\phi_1,\phi_2)^i .
\end{equation}
The conserved quantities associated with the three Killing directions
are
\begin{equation}
    \mathcal{Q}_i
    =
    \mathcal{G}_{ij}\dot z^j
    =
    (-E,L_1,L_2)_i ,
\end{equation}
and the geodesic normalisation condition becomes
\begin{equation}
\label{eq:velocities}
    \Lambda^2
    \left(
        \dot r^2+r^2\dot\theta^2
    \right)
    =
    -\kappa
    -
    \mathcal{G}^{ij}
    \mathcal{Q}_i\mathcal{Q}_j ,
\end{equation}
with $\kappa=1$ for timelike and $\kappa=0$ for null geodesics.

A particularly simple exact family consists of null or timelike
geodesics confined to the axis $\theta=0$, with
\begin{equation}
    L_1=0,
    \qquad
    L_2\neq0.
\end{equation}
Their radial motion satisfies
\begin{equation}
    \dot r^2
    =
    \left(
        E+j_2L_2r^2
    \right)^2
    -
    \kappa
    -
    \frac{L_2^2}{r^2}.
    \label{eq:axial-radial-geodesic}
\end{equation}
This equation can be integrated in terms of the Weierstrass elliptic
function,
\begin{equation}
    r^2(\lambda)
    =
    \frac{1}{(j_2L_2)^2}
    \Bigg[
    \wp\Bigg(
        \lambda-\lambda_\infty;
        \frac{4}{3}(j_2L_2)^2(E^2+3\kappa), 
        \frac{4}{27}(j_2L_2)^3
        \left(
            2E^3-18\kappa E+27j_2L_2^3
        \right)
    \Bigg)
    -
    \frac{2j_2L_2E}{3}
    \Bigg].
\label{eq:axial-geodesic-solution}
\end{equation}
Since
$\wp(u;g_2,g_3)\sim u^{-2}$ as $u\to0$
\cite{abramowitz1964handbook},
\begin{equation}
    r(\lambda)
    \underset{\lambda\to\lambda_\infty}{\sim}
    \frac{1}
    {|j_2L_2(\lambda-\lambda_\infty)|}.
    \label{eq:axial-geodesic-asymptotic}
\end{equation}
These causal geodesics therefore reach $r=\infty$ at finite affine
parameter.

Along the same axis, the curvature invariant is
\begin{equation}
    \mathcal{K}(r,0)
    =
    72j_2^4r^8
    -
    192(j_1^2+j_2^2)r^2,
\end{equation}
and hence
\begin{equation}
    \mathcal{K}(r,0)
    \sim
    72j_2^4r^8
    \qquad
    (r\to\infty).
\end{equation}
The spacetime is therefore causally geodesically incomplete, and the
divergence of the curvature invariant excludes a $C^2$ extension
through this asymptotic end. 

For completeness, neither of the two most symmetric parameter
restrictions removes the asymptotic singularities. First, setting one
of the swirl parameters to zero does not regularise the geometry. The
single-swirl limit is non-uniform: removing one contribution to
\eqref{eq:asymptotic-lambda} changes the dominant balance near the
corresponding angular endpoint and splits some of the regions above
into further subregions. Repeating the analysis yields thirteen rather
than eleven generic scaling regimes, with curvature divergences
persisting in additional branches.

The equal-swirl sector,
\begin{equation}
    j_1=j_2\equiv j\neq0,
\end{equation}
is likewise singular. In this case the enhanced plane-exchange symmetry
makes the two asymptotic corners equivalent, but does not alter their
curvature behaviour. In particular,
\begin{equation}
    \mathcal K(r,0)
    =
    72j^4r^8-384j^2r^2
    \sim72j^4r^8,
    \qquad r\to\infty,
\end{equation}
with the same leading divergence at $\theta=\pi/2$. The incomplete
causal geodesics discussed above also persist in this sector.
Consequently, neither suppressing one swirl nor imposing equal swirls
provides a route to an asymptotically regular background.

\section{The five-dimensional Schwarzschild--Tangherlini swirling black hole}
\label{sec4}

We now apply the double-axial Ehlers transformation to the
five-dimensional Schwarzschild--Tangherlini seed. The resulting
geometry carries two independent swirl parameters, $j_1$ and $j_2$,
associated with the two rotational planes. Its massless limit at fixed
$j_1$ and $j_2$ reproduces the doubly swirling background of
\autoref{sec3}, while setting either parameter to zero gives the
corresponding single-swirl black hole sector.

This also clarifies the relation with the single-swirl solution of
Ref.~\cite{Vigano:2026hqd}. The Schwarzschild--Tangherlini
configuration constructed there belongs to the present family, since
the black hole seed is reduced along its two axial Killing directions.
Its massless limit is therefore the single-swirl limit of the
axial-axial background derived here, rather than the
axial-translational background considered separately in that
reference. The distinction concerns the identification of the
background, not the validity of the axial-translational solution
itself.

Introducing
\begin{equation}
    \mu_1=\sin\theta,
    \qquad
    \mu_2=\cos\theta,
    \qquad
    0\leq\theta\leq\frac{\pi}{2},
\end{equation}
with $\phi_a\sim\phi_a+2\pi$, the metric reads
\begin{align}
\mathrm{ds}^2={}&
\frac{\mu_1^2r^2
\left(1+j_2^2\mu_1^2\mu_2^4r^6\right)}
{\Lambda^2}
\bm{\Theta}_1^2
+
\frac{\mu_2^2r^2
\left(1+j_1^2\mu_1^4\mu_2^2r^6\right)}
{\Lambda^2}
\bm{\Theta}_2^2
-
\frac{2j_1j_2\mu_1^4\mu_2^4r^8}
{\Lambda^2}
\bm{\Theta}_1\bm{\Theta}_2
\nonumber\\
&+
\Lambda^2
\left[
-\left(1-\frac{m}{r^2}\right)\diff t^2
+\left(1-\frac{m}{r^2}\right)^{-1}\diff r^2
+r^2\diff\theta^2
\right],
\label{eq:swirling-ST}
\end{align}
where
\begin{equation}
\begin{split}
\bm{\Theta}_1
&=
\diff\phi_1
+j_1(r^2-m)(3\mu_2^2-1)\diff t,
\\
\bm{\Theta}_2
&=
\diff\phi_2
+j_2(r^2-m)(3\mu_2^2-2)\diff t,
\end{split}
\label{eq:swirling-ST-theta}
\end{equation}
and
\begin{equation}
    \Lambda^2
    =
    1+\mu_1^2\mu_2^2r^6
    \left(
        j_1^2\mu_1^2+j_2^2\mu_2^2
    \right).
    \label{eq:swirling-ST-lambda}
\end{equation}

The hypersurface
\begin{equation}
    r_{\mathcal H}=\sqrt{m}
\end{equation}
is a regular Killing horizon. In the stationary frame naturally
inherited from the Schwarzschild--Tangherlini seed, its generator is
simply
\begin{equation}
    \bm{\chi}=\partial_t,
\end{equation}
so that
\begin{equation}
    \Omega_{\mathcal H}^{(1)}
    =
    \Omega_{\mathcal H}^{(2)}
    =
    0.
\end{equation}
This follows directly from the factor $r^2-m$ in
\eqref{eq:swirling-ST-theta}. As in the background geometry, a
constant redefinition of the stationary frame shifts absolute angular
velocities; the statement above therefore refers to the natural
seed-adapted frame used throughout this section. In this frame, the
black hole carries no additional horizon rotation relative to the
surrounding swirl.

The Komar angular momenta associated with both axial Killing fields
vanish. With the normalization of $\partial_t$ inherited from the
seed, the corresponding Komar mass is
\begin{equation}
    M_{\mathrm K}
    =
    \frac{3\pi m}{8G_5}.
\end{equation}
The horizon area and temperature are
\begin{equation}
    A_{\mathcal H}
    =
    2\pi^2m^{3/2},
    \qquad
    T_{\mathcal H}
    =
    \frac{1}{2\pi\sqrt{m}},
\end{equation}
and hence
\begin{equation}
    S
    =
    \frac{A_{\mathcal H}}{4G_5}
    =
    \frac{\pi^2m^{3/2}}{2G_5}.
\end{equation}
Neither the area nor the temperature depends on the swirl parameters.
For variations at fixed $j_1$ and $j_2$, one therefore recovers
\begin{equation}
    \diff M_{\mathrm K}
    =
    T_{\mathcal H}\diff S,
    \qquad
    M_{\mathrm K}
    =
    \frac{3}{2}T_{\mathcal H}S .
    \label{eq:ST-thermodynamics}
\end{equation}
Thus the external swirl deforms the intrinsic horizon geometry without
changing its area or temperature.

To see this deformation explicitly, consider a spatial cross-section
of the horizon. Its induced metric is
\begin{align}
\mathrm{ds}_{\mathcal H}^2={}
m\Lambda_{\mathcal H}^2\diff\theta^2
+
\frac{m\mu_1^2
\left(
1+j_2^2m^3\mu_1^2\mu_2^4
\right)}
{\Lambda_{\mathcal H}^2}
\diff\phi_1^2
+
\frac{m\mu_2^2
\left(
1+j_1^2m^3\mu_1^4\mu_2^2
\right)}
{\Lambda_{\mathcal H}^2}
\diff\phi_2^2
-
\frac{2j_1j_2m^4\mu_1^4\mu_2^4}
{\Lambda_{\mathcal H}^2}
\diff\phi_1\diff\phi_2,
\label{eq:ST-horizon-metric}
\end{align}
where
\begin{equation}
    \Lambda_{\mathcal H}^2
    =
    1+\mu_1^2\mu_2^2
    \left(
        \bar j_1^{\,2}\mu_1^2
        +
        \bar j_2^{\,2}\mu_2^2
    \right),
    \qquad
    \bar j_a\equiv j_a m^{3/2}.
\end{equation}

It is convenient to express its intrinsic Ricci scalar in terms of
\begin{equation}
    \bar R_{\mathcal H}
    \equiv
    mR_{\mathcal H},
    \qquad
    x\equiv\cos^2\theta.
\end{equation}
One finds
\begin{equation}
    \bar R_{\mathcal H}
    =
    \frac{2P(x)}
    {\Lambda_{\mathcal H}^6},
    \label{eq:ST-horizon-Ricci}
\end{equation}
where
\begin{align}
P(x)={}&
6(\bar j_1^{\,2}-\bar j_2^{\,2})^2x^6
-3(\bar j_1^{\,2}-\bar j_2^{\,2})
(7\bar j_1^{\,2}-5\bar j_2^{\,2})x^5
+
\left(
28\bar j_1^{\,4}
-37\bar j_1^{\,2}\bar j_2^{\,2}
+13\bar j_2^{\,4}
\right)x^4
\nonumber\\
&-
\left[
18\bar j_1^{\,4}
-14\bar j_1^{\,2}\bar j_2^{\,2}
+4\bar j_2^{\,4}
+27(\bar j_1^{\,2}-\bar j_2^{\,2})
\right]x^3
+
\left(
6\bar j_1^{\,4}
-\bar j_1^{\,2}\bar j_2^{\,2}
+45\bar j_1^{\,2}
-36\bar j_2^{\,2}
\right)x^2
\nonumber\\
&-
\left(
\bar j_1^{\,4}
+21\bar j_1^{\,2}
-12\bar j_2^{\,2}
\right)x
+3\bar j_1^{\,2}+3 .
\end{align}
For $\bar j_1=\bar j_2=0$, one recovers
$\bar R_{\mathcal H}=6$, as expected for the round
Schwarzschild--Tangherlini horizon. For nonzero swirls,
$R_{\mathcal H}$ acquires a nontrivial angular dependence, showing
that the horizon is intrinsically distorted while retaining its
original area. In the equal-swirl sector,
$\bar j_1=\bar j_2$, the deformation is symmetric under
$\theta\leftrightarrow\pi/2-\theta$, reflecting the plane-exchange
symmetry of the background.

\begin{figure}[t]
    \centering
    \includegraphics[]{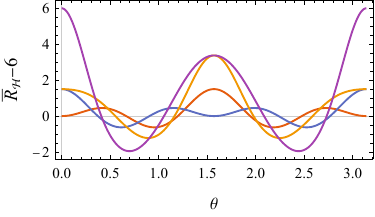}
    \caption{
    Angular dependence of the deviation
    $\bar R_{\mathcal H}-6$ of the dimensionless horizon Ricci scalar
    from its Schwarzschild--Tangherlini value, for
    $(\bar j_1,\bar j_2)=(0.5,0)$ (red),
    $(0,0.5)$ (blue),
    $(0.75,0.5)$ (yellow), and
    $(1,0.75)$ (purple).
    }
    \label{intrinsicricci}
\end{figure}

\section{Higher-dimensional swirling backgrounds}
\label{sec5}

We now examine how the preceding construction extends to arbitrary
dimension. The purpose of this section is not to establish a
nonexistence result for higher-dimensional swirling geometries, but
rather to determine which rotational sectors can be generated by the
particular Ehlers transformations considered here after reducing
vacuum General Relativity to three dimensions.

A $D$-dimensional reduction to three dimensions requires $D-3$
mutually commuting Killing vectors. In Minkowski spacetime, the
maximum number of mutually commuting rotations in orthogonal spatial
two-planes is
\begin{equation}
    q_{\rm max}
    =
    \left\lfloor\frac{D-1}{2}\right\rfloor .
\end{equation}
Therefore, a reduction performed entirely along axial directions
would require
\begin{equation}
    D-3
    \leq
    \left\lfloor\frac{D-1}{2}\right\rfloor ,
\end{equation}
which holds only for $D\leq5$. Five dimensions are thus the highest
dimension in which all Killing directions required by this
three-dimensional reduction can be chosen to be axial. This explains
the distinguished role of the doubly swirling geometry constructed in
\autoref{sec3}, but does not preclude other higher-dimensional
swirling solutions obtained by different methods.

More generally, write Minkowski spacetime as
\begin{equation}
    \mathrm{ds}_D^2
    =
    -\diff t^2
    +
    \sum_{m=1}^{q}
    \left(
        \diff r_m^2+r_m^2\diff\phi_m^2
    \right)
    +
    \sum_{\alpha=1}^{D-2q-1}
    \diff z_\alpha^2 .
    \label{eq:D-Minkowski}
\end{equation}
A purely spacelike reduction may use the $q$ axial directions together
with translational Killing vectors. The total number of mutually
commuting spacelike directions available in this form is
\begin{equation}
    q+(D-1-2q)=D-1-q .
\end{equation}
Requiring at least $D-3$ such directions gives
\begin{equation}
    q\leq2 .
\end{equation}
Thus, within this canonical purely spacelike reduction, two is the
largest number of axial sectors that can participate simultaneously.
For $D>5$, the remaining reduction directions must be translational.

We therefore consider the two-axial branch
\begin{equation}
\begin{split}
    \mathrm{ds}_D^2={}&
    -\diff t^2
    +\diff r_1^2+r_1^2\diff\phi_1^2
    +\diff r_2^2+r_2^2\diff\phi_2^2
    +\sum_{\alpha=1}^{D-5}\diff z_\alpha^2 ,
\end{split}
\end{equation}
and reduce along
\begin{equation}
    \left\{
        \partial_{\phi_1},
        \partial_{\phi_2},
        \partial_{z_1},
        \ldots,
        \partial_{z_{D-5}}
    \right\}.
\end{equation}
We denote by $j_1$ and $j_2$ the Ehlers parameters associated with the
two axial directions, and by $b_\alpha$ those associated with the
translations.

The different roles of these parameters follow from the Gram matrix
\begin{equation}
    (f_{ab})
    =
    \operatorname{diag}
    \left(
        r_1^2,r_2^2,1,\ldots,1
    \right),
    \qquad
    f=r_1^2r_2^2 .
\end{equation}
Although the translational Killing vectors have constant norm, they
belong to the same reduced sector as the two axial directions.
Consequently, their Ehlers parameters couple to the nontrivial
coordinate dependence of $f$ and generate genuine twists rather than
remaining spectators.

Introducing
\begin{equation}
    \rho=r_1r_2 ,
\end{equation}
the deformation function is
\begin{equation}
    \Lambda^2
    =
    1+\rho^2
    \left(
        j_1^2r_1^2+j_2^2r_2^2
        +\sum_{\alpha=1}^{D-5}b_\alpha^2
    \right),
    \label{eq:D-Lambda}
\end{equation}
while the reconstructed 1-forms are
\begin{equation}
    \Theta_1
    =
    \diff\phi_1
    +j_1(r_1^2-2r_2^2)\diff t,
    \qquad
    \Theta_2
    =
    \diff\phi_2
    +j_2(2r_1^2-r_2^2)\diff t,
\end{equation}
and
\begin{equation}
    \Xi_\alpha
    =
    \diff z_\alpha
    +
    b_\alpha(r_1^2-r_2^2)\diff t .
\end{equation}
The transformed vacuum metric takes the compact form
\begin{equation}
\begin{split}
\mathrm{ds}_D^2={}&
\Lambda^2
\left(
-\diff t^2+\diff r_1^2+\diff r_2^2
\right)
+r_1^2\Theta_1^2
+r_2^2\Theta_2^2
+\sum_{\alpha=1}^{D-5}\Xi_\alpha^2
\\
&-
\frac{\rho^2}{\Lambda^2}
\left(
j_1r_1^2\Theta_1
+j_2r_2^2\Theta_2
+\sum_{\alpha=1}^{D-5}
b_\alpha\Xi_\alpha
\right)^2 .
\label{eq:D-swirl}
\end{split}
\end{equation}
The parameters $j_1$ and $j_2$ generate the two rotational dragging
sectors discussed previously. By contrast, $b_\alpha$ produces a
position-dependent translational dragging,
\begin{equation}
    g_{tz_\alpha}\neq0,
\end{equation}
which cannot, in general, be removed by a constant boost.
While the Kaluza--Klein 1-form $\Xi_\alpha$ reverses its orientation across
$r_1=r_2$, the corresponding metric component $g_{t z_\alpha}$ does not, in
general, vanish or change sign there.

Six dimensions provide the first case in which such a translational
sector is unavoidable in the present purely spacelike construction.
The reduction uses
\begin{equation}
    \left\{
        \partial_{\phi_1},
        \partial_{\phi_2},
        \partial_z
    \right\},
\end{equation}
with Ehlers parameters $(j_1,j_2,b)$, and
\begin{equation}
    \Xi
    =
    \diff z+b(r_1^2-r_2^2)\diff t .
\end{equation}
For $b\neq0$, the resulting six-dimensional geometry is not the direct
product of the five-dimensional swirling background with a flat line.
It is therefore the first member of this family containing a genuinely
nontrivial translational Ehlers deformation.

For $D>6$, the translational parameters do not generate further
independent structures. They form a vector
\begin{equation}
    \bm{b}
    =
    (b_1,\ldots,b_{D-5})
\end{equation}
in the Euclidean space of translational Killing directions. Since the
Minkowski seed is invariant under orthogonal transformations among the
$z_\alpha$, one may align this vector with a single coordinate $z$.
Defining
\begin{equation}
    b^2
    =
    \sum_{\alpha=1}^{D-5}b_\alpha^2 ,
\end{equation}
the remaining $D-6$ translational directions become flat spectators.
Consequently, within this two-axial Ehlers family,
\begin{equation}
    \mathcal{M}^{\rm Ehl}_D
    \simeq
    \mathcal{M}^{\rm Ehl}_6
    \times
    \mathbb{R}^{D-6},
    \qquad
    D\geq6 .
    \label{eq:D-factorization}
\end{equation}
When $b=0$, this reduces to
\begin{equation}
    \mathcal{M}^{\rm Ehl}_D
    \simeq
    \mathcal{M}^{\rm 2-s}_5
    \times
    \mathbb{R}^{D-5}.
\end{equation}

It is natural to ask whether one can instead activate three independent
rotational planes. For example, seven-dimensional Minkowski spacetime
may be written as
\begin{equation}
    \mathrm{ds}_7^2
    =
    -\diff t^2
    +
    \sum_{m=1}^{3}
    \left(
        \diff r_m^2+r_m^2\diff\phi_m^2
    \right).
\end{equation}
A reduction to three dimensions requires four commuting Killing
directions. Since the three rotational planes exhaust the six spatial
coordinates in this parametrisation, a reduction retaining all three
axial vectors must include $\partial_t$,
\begin{equation}
    \left\{
        \partial_t,
        \partial_{\phi_1},
        \partial_{\phi_2},
        \partial_{\phi_3}
    \right\}.
\end{equation}
This belongs to the timelike--spacelike realisation of the sigma model
and is subject to the positivity condition
\eqref{eq:EhlersPositivity}.

For the Minkowski seed,
\begin{equation}
    f_{ab}
    =
    \operatorname{diag}
    \left(
        -1,r_1^2,r_2^2,r_3^2
    \right),
    \qquad
    f=-r_1^2r_2^2r_3^2 ,
\end{equation}
and the seed twist potentials vanish. Writing the Ehlers parameters as
$(j_t,j_1,j_2,j_3)$, one obtains
\begin{equation}
\begin{split}
    f\Pi
    =
    1
    +j_t^2r_1^2r_2^2r_3^2
    -
    r_1^2r_2^2r_3^2
    \left(
        j_1^2r_1^2
        +j_2^2r_2^2
        +j_3^2r_3^2
    \right).
    \label{eq:three-plane-positivity}
\end{split}
\end{equation}
The transformation remains in the same Lorentzian branch only while
$f\Pi>0$. Any nonzero axial Ehlers parameter eventually drives
\eqref{eq:three-plane-positivity} negative along suitable large-radius
directions. Thus the present Ehlers transformation cannot generate a
three-plane swirling geometry that covers the full domain
$r_m\in[0,\infty)$ while remaining on the same nondegenerate Lorentzian
branch.

This obstruction is specific to the solution-generating construction
used here. It does not imply that globally defined three-plane, or more
general higher-dimensional, swirling spacetimes do not exist. Such
geometries may require a different hidden symmetry transformation, a
different reduction scheme, or an independent construction not based
on the present Ehlers family.

Finally, the translational sector also clarifies the relation with the
axial-translational construction of Ref.~\cite{Vigano:2026hqd}. In
the full coset formulation, an axial-translational reduction admits
both an axial Ehlers parameter $j$ and a translational one $b$. The
restricted Ernst representation used there retains only the axial
sector. The additional $b$ deformation exhibited above is therefore
another feature that becomes visible only in the full sigma model
description.

\section{Closing remarks}
\label{sec6}

We have constructed a class of higher-dimensional swirling vacuum
geometries using the hidden symmetries of General Relativity reduced
to three dimensions. The key step is to work directly with the full
coset-space formulation. In five dimensions, reduction along the two
independent axial Killing vectors leads to an
$SL(3,\mathbb{R})/SO(3)$ sigma model and allows two independent
Ehlers parameters to be activated. The resulting background is
intrinsically five-dimensional and is inequivalent to the direct product
of the four-dimensional swirling spacetime with a flat spectator
direction \cite{Vigano:2026hqd}. This distinction persists even when one of the two swirl
parameters is set to zero.

Applying the same transformation to the Schwarzschild--Tangherlini
seed yields a doubly swirling black hole whose massless limit
consistently reproduces the corresponding axial-axial background.
The external swirl deforms the intrinsic horizon geometry while
leaving its area and temperature unchanged. We have also given the
general doubly swirling Myers--Perry solution, in which two intrinsic
black hole rotation parameters coexist with the two background swirls;
its detailed physical analysis is deferred to a companion work
\cite{toappear}.

A central result of this work is the multidirectional character of the
five-dimensional asymptotic geometry. The limit $r\to\infty$ is not
unique unless the simultaneous angular scaling is specified.
Correlated approaches to the two angular endpoints reveal asymptotic
regimes that are invisible in the usual fixed-angle limit. Although
the Kretschmann invariant decays at fixed interior angle, it diverges
along particular approaches to the asymptotic corners.

These directional divergences correspond to genuine singular
asymptotic ends. Along the axis $\theta=0$, exact null and timelike
geodesics with $L_1=0$ and $j_2L_2\neq0$ satisfy
\begin{equation}
    r(\lambda)
    \sim
    \frac{1}
    {|j_2L_2(\lambda-\lambda_\infty)|},
\end{equation}
and therefore reach $r=\infty$ in finite affine parameter. Along the
same trajectories,
\begin{equation}
    \mathcal K
    \sim
    72j_2^4r^8 .
\end{equation}
The spacetime is thus causally geodesically incomplete, and the
curvature divergence excludes a $C^2$ extension through this end.
The singular behaviour is therefore not an artefact of the coordinate
description or of considering arbitrary asymptotic paths.

Neither of the simplest parameter restrictions removes this feature.
The single-swirl limit is non-uniform and leads to a refined
classification with thirteen rather than eleven generic scaling
regimes, with curvature divergences persisting in additional branches.
The equal-swirl sector $j_1=j_2$ is also singular: the enhanced
plane-exchange symmetry makes the two asymptotic corners equivalent,
but does not change their divergent curvature behaviour. Thus neither
suppressing one swirl nor imposing equal swirls regularises the
background.

The higher-dimensional analysis should be interpreted as a statement
about the reach of the present solution-generating method, rather than
as a nonexistence result. Within the Ehlers family studied here, five
dimensions are the highest dimension in which the reduction to three
dimensions can be performed entirely along axial Killing vectors.
For $D>5$, a purely spacelike reduction retaining two axial sectors
necessarily involves translational Killing directions. Their Ehlers
parameters generate position-dependent translational dragging, with
the first genuinely new case appearing in six dimensions. For
$D>6$, these translational parameters reduce, by rotational symmetry
among the flat directions, to a single effective deformation, while
the remaining directions are spectators.

Activating three independent rotational planes within the same
framework requires, already in seven dimensions, a reduction
containing the timelike Killing vector. For the Ehlers transformation
used here, the corresponding positivity condition fails at sufficiently
large radius whenever an axial Ehlers parameter is nonzero. The
construction therefore does not yield a globally defined three-plane
swirling geometry over the full radial domain. This obstruction is
specific to the present reduction and transformation: it does not
exclude the existence of higher-dimensional swirling spacetimes with
three or more rotational sectors obtained by different
solution-generating techniques or by direct construction.

The curvature-singular asymptotics suggest several directions for
future work. One possibility is to investigate whether additional
fields can modify the asymptotic balance responsible for the
divergences. External electromagnetic fields are a natural candidate,
although presently available higher-dimensional magnetising
transformations impose restrictions that make the doubly swirling
sector difficult to access directly. More generally, it would be
interesting to determine whether alternative hidden symmetry
transformations or different reduction schemes can generate
higher-dimensional swirling backgrounds with improved asymptotic
behaviour and with more than two independent rotational sectors.

\acknowledgments
J.B. acknowledges the support of ANID through grant FOVI250117 and the hospitality of the Institute of Mathematics, Czech Academy of Sciences (RVO 67985840). A.C. is partially supported by FONDECYT grant 1250318. C. H. is  supported by 
CIDMA (\url{https://ror.org/05pm2mw36}) under the Portuguese Foundation for Science and Technology (FCT, \url{https://ror.org/ 00snfqn58}), Grants UID/04106/2025 (\url{https://doi.org/10.54499/UID/ 04106/2025}) and UID/PRR/04106/2025    
(\url{https://doi.org/10.54499/UID/PRR/ 04106/2025}), as well as the Horizon Europe staff exchange (SE) programme HORIZON-MSCA2021-SE-01 Grant No.\ NewFunFiCO-101086251.

\appendix

\appendix

\section{Kretschmann invariant of the five-dimensional swirling background}
\label{app:kretschmann}

For completeness, we give here the exact Kretschmann invariant used in
the asymptotic analysis of \autoref{subsec:background-asymptotics}.
Introducing the double-polar coordinates
\begin{equation}
    r_1=r\sin\theta,
    \qquad
    r_2=r\cos\theta,
\end{equation}
it is convenient to decompose
\begin{equation}
    \mathcal K
    \equiv
    R_{\mu\nu\rho\sigma}R^{\mu\nu\rho\sigma}
    =
    \mathcal K_1+\mathcal K_2+\mathcal K_{12}.
\end{equation}
The three contributions are
\begin{align}
\mathcal{K}_1={}&
\frac{24j_1^2}
{\left(
1+j_1^2r_1^4r_2^2+j_2^2r_1^2r_2^4
\right)^6}
\Big[
r_1^{12}r_2^4
\left(
3r_1^4+4r_1^2r_2^2+8r_2^4
\right)j_1^6
-2r_1^8r_2^2
\left(
5r_1^4+24r_1^2r_2^2+60r_2^4
\right)j_1^4
\nonumber\\
&\hspace{2cm}
+r_1^4
\left(
3r_1^4+68r_1^2r_2^2+120r_2^4
\right)j_1^2
-8r_1^2-8r_2^2
\Big],
\label{K1}
\end{align}
\begin{align}
\mathcal{K}_2={}&
\frac{24j_2^2}
{\left(
1+j_1^2r_1^4r_2^2+j_2^2r_1^2r_2^4
\right)^6}
\Big[
r_1^4r_2^{12}
\left(
8r_1^4+4r_1^2r_2^2+3r_2^4
\right)j_2^6
-2r_1^2r_2^8
\left(
60r_1^4+24r_1^2r_2^2+5r_2^4
\right)j_2^4
\nonumber\\
&\hspace{2cm}
+r_2^4
\left(
120r_1^4+68r_1^2r_2^2+3r_2^4
\right)j_2^2
-8r_1^2-8r_2^2
\Big],
\label{K2}
\end{align}
and
\begin{align}
\mathcal{K}_{12}={}&
\frac{24j_1^2j_2^2r_1^2r_2^2}
{\left(
1+j_1^2r_1^4r_2^2+j_2^2r_1^2r_2^4
\right)^6}
\Big[
2r_1^8r_2^4
\left(
7r_1^4+5r_1^2r_2^2+4r_2^4
\right)j_1^4
+2r_1^4r_2^8
\left(
4r_1^4+5r_1^2r_2^2+7r_2^4
\right)j_2^4
\nonumber\\
&\hspace{1.2cm}
+\Big[
r_1^6r_2^6
\left(
27r_1^4+44r_1^2r_2^2+27r_2^4
\right)j_2^2
-2r_1^4r_2^2
\left(
25r_1^4+118r_1^2r_2^2+40r_2^4
\right)
\Big]j_1^2
\nonumber\\
&\hspace{1.2cm}
-2r_1^2r_2^4
\left(
40r_1^4+118r_1^2r_2^2+25r_2^4
\right)j_2^2
+40r_1^4+178r_1^2r_2^2+40r_2^4
\Big].
\label{K12}
\end{align}
Their common denominator is $\Lambda^{12}$, with
\begin{equation}
    \Lambda^2
    =
    1+j_1^2r_1^4r_2^2+j_2^2r_1^2r_2^4 .
\end{equation}
The asymptotic expansions of these exact expressions in the correlated
radial--angular limits are summarised in
\autoref{tab:kretschmann-eleven-regions}.

\section{The general five-dimensional swirling Myers--Perry geometry}
\label{app:swirlingMP}

We collect here the general doubly swirling Myers--Perry solution obtained by applying the axial-axial Ehlers transformation to the five-dimensional Myers--Perry seed.\footnote{A Mathematica notebook containing the solution is available as an ancillary file in the arXiv source folder.} The geometry depends on the mass parameter $\mu$, the two intrinsic rotation parameters $a_1,a_2$, and the two background swirl parameters $j_1,j_2$. Its detailed physical analysis will be presented elsewhere \cite{toappear}.

The doubly rotating Myers--Perry metric is
\begin{equation}
\begin{split}
\mathrm{ds}_{\rm MP}^2={}
-\diff t^2
+
\frac{\mu}{\Sigma}
\left(
\diff t
+a_1\mu_1^2\diff\phi_1
+a_2\mu_2^2\diff\phi_2
\right)^2
+
\frac{\Sigma r^2}{\Delta}\diff r^2
+\Sigma\diff\theta^2
+
\sum_{i=1}^{2}
\left(
r^2+a_i^2
\right)\mu_i^2\diff\phi_i^2 ,
\end{split}
\label{eq:MP-seed}
\end{equation}
where
\begin{equation}
    \mu_1=\sin\theta,
    \qquad
    \mu_2=\cos\theta,
\end{equation}
and
\begin{equation}
    \Sigma
    =
    r^2+a_1^2\mu_2^2+a_2^2\mu_1^2,
    \qquad
    \Delta
    =
    (r^2+a_1^2)(r^2+a_2^2)-\mu r^2 .
\end{equation}

We reduce along
\begin{equation}
    x^a=(\phi_1,\phi_2)^a .
\end{equation}
The corresponding Gram matrix is
\begin{equation}
\begin{split}
f_{11}&=
\mu_1^2(r^2+a_1^2)
+
\frac{\mu a_1^2\mu_1^4}{\Sigma},
\\
f_{22}&=
\mu_2^2(r^2+a_2^2)
+
\frac{\mu a_2^2\mu_2^4}{\Sigma},
\\
f_{12}&=
\frac{\mu a_1a_2\mu_1^2\mu_2^2}{\Sigma},
\end{split}
\label{eq:MP-Gram}
\end{equation}
with
\begin{equation}
\begin{split}
f\equiv\det(f_{ab})
=
\mu_1^2\mu_2^2
\Bigg[
(r^2+a_1^2)(r^2+a_2^2)
+
\frac{\mu}{\Sigma}
\left(
a_1^2\mu_1^2(r^2+a_2^2)
+
a_2^2\mu_2^2(r^2+a_1^2)
\right)
\Bigg].
\end{split}
\label{eq:MP-f}
\end{equation}
The nonvanishing time-angular components of the seed are
\begin{equation}
    g_{ta}
    =
    \frac{\mu a_a\mu_a^2}{\Sigma},
\end{equation}
and a convenient choice of twist potentials is
\begin{equation}
    \chi_1
    =
    -\frac{\mu a_1(r^2+a_2^2)\mu_1^4}{\Sigma},
    \qquad
    \chi_2
    =
    \frac{\mu a_2(r^2+a_1^2)\mu_2^4}{\Sigma}.
    \label{eq:MP-twists}
\end{equation}

The transformed metric can be written as
\begin{equation}
    \mathrm{ds}^2
    =
    f'_{ab}
    \left(
        \diff x^a+\mathcal A'^a
    \right)
    \left(
        \diff x^b+\mathcal A'^b
    \right)
    +
    \frac{\Lambda^2}{f}
    \diff\tilde s_3^2 ,
    \label{eq:swirling-MP}
\end{equation}
where
\begin{equation}
\diff\tilde {\rm s}_3^2
=
f
\left[
\left(
-1+\frac{\mu}{\Sigma}
-g_{ta}g_{tb}f^{ab}
\right)\diff t^2
+
\frac{\Sigma r^2}{\Delta}\diff r^2
+
\Sigma\diff\theta^2
\right].
\label{eq:MP-base}
\end{equation}

For
\begin{equation}
    \bm j=(j_1,j_2)^T,
\end{equation}
the deformation function is most compactly written as
\begin{equation}
    \Lambda^2
    =
    f\,j^af_{ab}j^b
    +
    \left(
        1+j^a\chi_a
    \right)^2 .
    \label{eq:MP-Lambda-compact}
\end{equation}
Explicitly,
\begin{align}
\Lambda^2={}&
f
\Bigg[
j_1^2\mu_1^2
\left(
r^2+a_1^2+
\frac{\mu a_1^2\mu_1^2}{\Sigma}
\right)
+
\frac{2j_1j_2\mu a_1a_2\mu_1^2\mu_2^2}{\Sigma}
+j_2^2\mu_2^2
\left(
r^2+a_2^2+
\frac{\mu a_2^2\mu_2^2}{\Sigma}
\right)
\Bigg]
\nonumber\\
&+
\left[
1
-\frac{\mu a_1j_1(r^2+a_2^2)\mu_1^4}{\Sigma}
+\frac{\mu a_2j_2(r^2+a_1^2)\mu_2^4}{\Sigma}
\right]^2 .
\label{eq:MP-Lambda}
\end{align}

Rather than expanding the transformed Gram matrix, it is convenient to
use directly the Ehlers formulas of \autoref{sec2}. The transformed
twist potentials are
\begin{equation}
    \chi'_a
    =
    \frac{
        f f_{ab}j^b
        +
        \chi_a(1+j^b\chi_b)
    }
    {\Lambda^2},
    \label{eq:MP-transformed-twists}
\end{equation}
and
\begin{equation}
    f'_{ab}
    =
    f_{ab}
    +
    \frac{1}{f}\chi_a\chi_b
    -
    \frac{\Lambda^2}{f}\chi'_a\chi'_b .
    \label{eq:MP-transformed-Gram}
\end{equation}
Equations
\eqref{eq:MP-Gram}--\eqref{eq:MP-transformed-Gram} therefore give the
transformed scalar sector explicitly.

The reconstructed Kaluza--Klein 1-forms have only a stationary
component and may be written as
\begin{equation}
    \mathcal A'^a
    =
    \frac{\mu_1^2\mu_2^2}{f\Sigma}
    \sum_{n=0}^{3}
    \alpha_n^{(a)}(r)
    P_n(\cos2\theta)\,\diff t,
    \label{eq:MP-KK}
\end{equation}
where $P_n$ are Legendre polynomials and
\begin{equation}
    \alpha_m^{(a)}(r)
    =
    \sum_{n=m}^{3}
    \frac{(2m+1)(n!)^2}
    {(n-m)!(n+m+1)!}
    \,
    l_n^{(a)}(r).
    \label{eq:MP-alpha}
\end{equation}

For the first rotational sector,
\begin{align}
l_0^{(1)}={}&
(r^2+a_2^2)(j_1a_1\mu-1)
\Bigg[
j_1(r^2+a_1^2)
(r^2-a_1^2-2a_2^2)
(r^2+a_2^2)
-a_1
\left[
1+j_1a_1(-r^2+a_1^2+2a_2^2)
\right]\mu
+j_1a_1^2\mu^2
\Bigg],
\nonumber\\
l_1^{(1)}={}&
-j_1\Delta(j_1a_1\mu-1)
\left[
3r^4+6r^2a_2^2
+a_1^2(a_1^2+2a_2^2-\mu)
\right],
\nonumber\\
l_2^{(1)}={}&
\Delta
\Bigg[
j_1(a_1^2-a_2^2)
(3r^2+a_1^2+2a_2^2)
+j_1a_1a_2(j_2a_1-2j_1a_2)\mu^2
+
\left[
3j_1^2a_1(r^2+a_2^2)^2
-j_2a_1a_2
-j_1(a_1^2-2a_2^2)
\right]\mu
\Bigg],
\nonumber\\
l_3^{(1)}={}&
j_1\mu\Delta
\Bigg[
j_2a_2
\left[
(r^2+a_1^2)^2-a_1^2\mu
\right]
-j_1a_1
\left[
(r^2+a_2^2)^2-a_2^2\mu
\right]
\Bigg].
\label{eq:l-first}
\end{align}

For the second rotational sector,
\begin{align}
l_0^{(2)}={}&
a_2\mu
\left\{
(r^2+a_1^2)
\left[
1+j_1a_1(r^2+a_2^2)
\right]
-j_1r^2a_1\mu
\right\}
+
j_2(r^2+a_2^2)
\Bigg[
-2(r^2+a_1^2)
(r^2-a_1^2-2a_2^2)
(r^2+a_2^2)
\nonumber\\
&\hspace{0.8cm}
+a_1(r^2-a_1^2-2a_2^2)
\left[
-2a_1+j_1(r^2+a_1^2)(r^2+a_2^2)
\right]\mu
-a_1^2
\left[
1+j_1a_1(-r^2+a_1^2+2a_2^2)
\right]\mu^2
+j_1a_1^3\mu^3
\Bigg],
\nonumber\\
l_1^{(2)}={}&
-\Delta
\Bigg[
2j_1a_1a_2\mu
+j_2
\Big[
-2a_1^4+3a_2^4
+j_1a_1^5\mu
+j_1a_1^3(2a_2^2-\mu)\mu
\nonumber\\
&\hspace{2cm}
-a_1^2(4a_2^2+\mu)
+3r^4(j_1a_1\mu-1)
+6r^2a_2^2(j_1a_1\mu-1)
\Big]
\Bigg],
\nonumber\\
l_2^{(2)}={}&
\Delta
\Bigg[
j_2(a_1^2-a_2^2)
(3r^2+2a_1^2+a_2^2)
+j_2(-2a_1^2+a_2^2)\mu
\nonumber\\
&\hspace{1.5cm}
+j_1a_1
\left[
a_2+3j_2(r^2+a_2^2)^2
\right]\mu
+j_2a_1a_2(j_2a_1-2j_1a_2)\mu^2
\Bigg],
\nonumber\\
l_3^{(2)}={}&
j_2\mu\Delta
\Bigg[
j_2a_2
\left[
(r^2+a_1^2)^2-a_1^2\mu
\right]
-j_1a_1
\left[
(r^2+a_2^2)^2-a_2^2\mu
\right]
\Bigg].
\label{eq:l-second}
\end{align}

The solution contains the expected limiting geometries. For
\begin{equation}
    j_1=j_2=0,
\end{equation}
it reduces to the doubly rotating Myers--Perry spacetime, whereas
\begin{equation}
    a_1=a_2=0,
\end{equation}
gives, after constant shifts of the angular coordinates, the doubly swirling Schwarzschild--Tangherlini solution of \autoref{sec4}.
 
\bibliography{biblio}

\begin{thebibliography}{35}%
\makeatletter
\providecommand \@ifxundefined [1]{%
 \@ifx{#1\undefined}
}%
\providecommand \@ifnum [1]{%
 \ifnum #1\expandafter \@firstoftwo
 \else \expandafter \@secondoftwo
 \fi
}%
\providecommand \@ifx [1]{%
 \ifx #1\expandafter \@firstoftwo
 \else \expandafter \@secondoftwo
 \fi
}%
\providecommand \natexlab [1]{#1}%
\providecommand \enquote  [1]{``#1''}%
\providecommand \bibnamefont  [1]{#1}%
\providecommand \bibfnamefont [1]{#1}%
\providecommand \citenamefont [1]{#1}%
\providecommand \href@noop [0]{\@secondoftwo}%
\providecommand \href [0]{\begingroup \@sanitize@url \@href}%
\providecommand \@href[1]{\@@startlink{#1}\@@href}%
\providecommand \@@href[1]{\endgroup#1\@@endlink}%
\providecommand \@sanitize@url [0]{\catcode `\\12\catcode `\$12\catcode `\&12\catcode `\#12\catcode `\^12\catcode `\_12\catcode `\%12\relax}%
\providecommand \@@startlink[1]{}%
\providecommand \@@endlink[0]{}%
\providecommand \url  [0]{\begingroup\@sanitize@url \@url }%
\providecommand \@url [1]{\endgroup\@href {#1}{\urlprefix }}%
\providecommand \urlprefix  [0]{URL }%
\providecommand \Eprint [0]{\href }%
\providecommand \doibase [0]{https://doi.org/}%
\providecommand \selectlanguage [0]{\@gobble}%
\providecommand \bibinfo  [0]{\@secondoftwo}%
\providecommand \bibfield  [0]{\@secondoftwo}%
\providecommand \translation [1]{[#1]}%
\providecommand \BibitemOpen [0]{}%
\providecommand \bibitemStop [0]{}%
\providecommand \bibitemNoStop [0]{.\EOS\space}%
\providecommand \EOS [0]{\spacefactor3000\relax}%
\providecommand \BibitemShut  [1]{\csname bibitem#1\endcsname}%
\let\auto@bib@innerbib\@empty
\bibitem [{\citenamefont {Ehlers}(1957)}]{Ehlers:1957zz}%
  \BibitemOpen
  \bibfield  {author} {\bibinfo {author} {\bibfnamefont {J.}~\bibnamefont {Ehlers}},\ }\emph {\bibinfo {title} {Konstruktionen und Charakterisierung von L{\"o}sungen der Einsteinschen Gravitationsfeldgleichungen}},\ \href@noop {} {Ph.D. thesis},\ \bibinfo  {school} {Universit{\"a}t Hamburg} (\bibinfo {year} {1957})\BibitemShut {NoStop}%
\bibitem [{\citenamefont {Maison}(1979)}]{Maison:1979kt}%
  \BibitemOpen
  \bibfield  {author} {\bibinfo {author} {\bibfnamefont {D.}~\bibnamefont {Maison}},\ }\bibfield  {title} {\bibinfo {title} {Ehlers-harrison-type transformations for jordan's extended theory of gravitation},\ }\href {https://doi.org/10.1007/BF00756907} {\bibfield  {journal} {\bibinfo  {journal} {General Relativity and Gravitation}\ }\textbf {\bibinfo {volume} {10}},\ \bibinfo {pages} {717} (\bibinfo {year} {1979})}\BibitemShut {NoStop}%
\bibitem [{\citenamefont {Ernst}(1968{\natexlab{a}})}]{Ernst:1967by}%
  \BibitemOpen
  \bibfield  {author} {\bibinfo {author} {\bibfnamefont {F.~J.}\ \bibnamefont {Ernst}},\ }\bibfield  {title} {\bibinfo {title} {{New Formulation of the Axially Symmetric Gravitational Field Problem. II}},\ }\href {https://doi.org/10.1103/PhysRev.168.1415} {\bibfield  {journal} {\bibinfo  {journal} {Phys. Rev.}\ }\textbf {\bibinfo {volume} {168}},\ \bibinfo {pages} {1415} (\bibinfo {year} {1968}{\natexlab{a}})}\BibitemShut {NoStop}%
\bibitem [{\citenamefont {Ernst}(1968{\natexlab{b}})}]{Ernst:1967wx}%
  \BibitemOpen
  \bibfield  {author} {\bibinfo {author} {\bibfnamefont {F.~J.}\ \bibnamefont {Ernst}},\ }\bibfield  {title} {\bibinfo {title} {{New formulation of the axially symmetric gravitational field problem}},\ }\href {https://doi.org/10.1103/PhysRev.167.1175} {\bibfield  {journal} {\bibinfo  {journal} {Phys. Rev.}\ }\textbf {\bibinfo {volume} {167}},\ \bibinfo {pages} {1175} (\bibinfo {year} {1968}{\natexlab{b}})}\BibitemShut {NoStop}%
\bibitem [{\citenamefont {Kinnersley}(1973)}]{Kinnersley1973}%
  \BibitemOpen
  \bibfield  {author} {\bibinfo {author} {\bibfnamefont {W.}~\bibnamefont {Kinnersley}},\ }\bibfield  {title} {\bibinfo {title} {Generation of stationary einstein--maxwell fields},\ }\href {https://doi.org/10.1063/1.1666368} {\bibfield  {journal} {\bibinfo  {journal} {Journal of Mathematical Physics}\ }\textbf {\bibinfo {volume} {14}},\ \bibinfo {pages} {651} (\bibinfo {year} {1973})}\BibitemShut {NoStop}%
\bibitem [{\citenamefont {Kinnersley}(1977)}]{Kinnersley:1977}%
  \BibitemOpen
  \bibfield  {author} {\bibinfo {author} {\bibfnamefont {W.}~\bibnamefont {Kinnersley}},\ }\bibfield  {title} {\bibinfo {title} {{Symmetries of the Stationary Einstein-Maxwell Field Equations. 1.}},\ }\href {https://doi.org/10.1063/1.523458} {\bibfield  {journal} {\bibinfo  {journal} {J. Math. Phys.}\ }\textbf {\bibinfo {volume} {18}},\ \bibinfo {pages} {1529} (\bibinfo {year} {1977})}\BibitemShut {NoStop}%
\bibitem [{\citenamefont {Kinnersley}\ and\ \citenamefont {Chitre}(1977)}]{KinnersleyChitre:1977}%
  \BibitemOpen
  \bibfield  {author} {\bibinfo {author} {\bibfnamefont {W.}~\bibnamefont {Kinnersley}}\ and\ \bibinfo {author} {\bibfnamefont {D.~M.}\ \bibnamefont {Chitre}},\ }\bibfield  {title} {\bibinfo {title} {Symmetries of the stationary einstein--maxwell field equations. ii},\ }\href {https://doi.org/10.1063/1.523459} {\bibfield  {journal} {\bibinfo  {journal} {J. Math. Phys.}\ }\textbf {\bibinfo {volume} {18}},\ \bibinfo {pages} {1538} (\bibinfo {year} {1977})}\BibitemShut {NoStop}%
\bibitem [{\citenamefont {Kinnersley}\ and\ \citenamefont {Chitre}(1978{\natexlab{a}})}]{KinnersleyChitre:1978a}%
  \BibitemOpen
  \bibfield  {author} {\bibinfo {author} {\bibfnamefont {W.}~\bibnamefont {Kinnersley}}\ and\ \bibinfo {author} {\bibfnamefont {D.~M.}\ \bibnamefont {Chitre}},\ }\bibfield  {title} {\bibinfo {title} {Symmetries of the stationary einstein--maxwell field equations. iii},\ }\href {https://doi.org/10.1063/1.523912} {\bibfield  {journal} {\bibinfo  {journal} {J. Math. Phys.}\ }\textbf {\bibinfo {volume} {19}},\ \bibinfo {pages} {1926} (\bibinfo {year} {1978}{\natexlab{a}})}\BibitemShut {NoStop}%
\bibitem [{\citenamefont {Kinnersley}\ and\ \citenamefont {Chitre}(1978{\natexlab{b}})}]{KinnersleyChitre:1978b}%
  \BibitemOpen
  \bibfield  {author} {\bibinfo {author} {\bibfnamefont {W.}~\bibnamefont {Kinnersley}}\ and\ \bibinfo {author} {\bibfnamefont {D.~M.}\ \bibnamefont {Chitre}},\ }\bibfield  {title} {\bibinfo {title} {Symmetries of the stationary einstein--maxwell field equations. iv. transformations which preserve asymptotic flatness},\ }\href {https://doi.org/10.1063/1.523580} {\bibfield  {journal} {\bibinfo  {journal} {J. Math. Phys.}\ }\textbf {\bibinfo {volume} {19}},\ \bibinfo {pages} {2037} (\bibinfo {year} {1978}{\natexlab{b}})}\BibitemShut {NoStop}%
\bibitem [{\citenamefont {Geroch}(1971)}]{Geroch:1970nt}%
  \BibitemOpen
  \bibfield  {author} {\bibinfo {author} {\bibfnamefont {R.~P.}\ \bibnamefont {Geroch}},\ }\bibfield  {title} {\bibinfo {title} {{A Method for generating solutions of Einstein's equations}},\ }\href {https://doi.org/10.1063/1.1665681} {\bibfield  {journal} {\bibinfo  {journal} {J. Math. Phys.}\ }\textbf {\bibinfo {volume} {12}},\ \bibinfo {pages} {918} (\bibinfo {year} {1971})}\BibitemShut {NoStop}%
\bibitem [{\citenamefont {Geroch}(1972)}]{Geroch:1972}%
  \BibitemOpen
  \bibfield  {author} {\bibinfo {author} {\bibfnamefont {R.}~\bibnamefont {Geroch}},\ }\bibfield  {title} {\bibinfo {title} {A method for generating new solutions of einstein's equation. ii},\ }\href {https://doi.org/10.1063/1.1665990} {\bibfield  {journal} {\bibinfo  {journal} {J. Math. Phys.}\ }\textbf {\bibinfo {volume} {13}},\ \bibinfo {pages} {394} (\bibinfo {year} {1972})}\BibitemShut {NoStop}%
\bibitem [{\citenamefont {Belinski}\ and\ \citenamefont {Zakharov}(1978)}]{Belinski:1978}%
  \BibitemOpen
  \bibfield  {author} {\bibinfo {author} {\bibfnamefont {V.~A.}\ \bibnamefont {Belinski}}\ and\ \bibinfo {author} {\bibfnamefont {V.~E.}\ \bibnamefont {Zakharov}},\ }\bibfield  {title} {\bibinfo {title} {Integration of the einstein equations by the inverse scattering problem},\ }\href@noop {} {\bibfield  {journal} {\bibinfo  {journal} {Soviet Physics JETP}\ }\textbf {\bibinfo {volume} {48}},\ \bibinfo {pages} {985} (\bibinfo {year} {1978})},\ \bibinfo {note} {translated from Zh. Eksp. Teor. Fiz. 75, 1953--1971 (1978)}\BibitemShut {NoStop}%
\bibitem [{\citenamefont {Belinski}\ and\ \citenamefont {Zakharov}(1979)}]{Belinski:1979}%
  \BibitemOpen
  \bibfield  {author} {\bibinfo {author} {\bibfnamefont {V.~A.}\ \bibnamefont {Belinski}}\ and\ \bibinfo {author} {\bibfnamefont {V.~E.}\ \bibnamefont {Zakharov}},\ }\bibfield  {title} {\bibinfo {title} {Stationary gravitational solitons with axial symmetry},\ }\href@noop {} {\bibfield  {journal} {\bibinfo  {journal} {Soviet Physics JETP}\ }\textbf {\bibinfo {volume} {50}},\ \bibinfo {pages} {1} (\bibinfo {year} {1979})},\ \bibinfo {note} {translated from Zh. Eksp. Teor. Fiz. 77, 3--19 (1979)}\BibitemShut {NoStop}%
\bibitem [{\citenamefont {Taub}(1951)}]{Taub:1950ez}%
  \BibitemOpen
  \bibfield  {author} {\bibinfo {author} {\bibfnamefont {A.~H.}\ \bibnamefont {Taub}},\ }\bibfield  {title} {\bibinfo {title} {{Empty space-times admitting a three parameter group of motions}},\ }\href {https://doi.org/10.2307/1969567} {\bibfield  {journal} {\bibinfo  {journal} {Annals Math.}\ }\textbf {\bibinfo {volume} {53}},\ \bibinfo {pages} {472} (\bibinfo {year} {1951})}\BibitemShut {NoStop}%
\bibitem [{\citenamefont {Newman}\ \emph {et~al.}(1963)\citenamefont {Newman}, \citenamefont {Tamburino},\ and\ \citenamefont {Unti}}]{Newman:1963yy}%
  \BibitemOpen
  \bibfield  {author} {\bibinfo {author} {\bibfnamefont {E.}~\bibnamefont {Newman}}, \bibinfo {author} {\bibfnamefont {L.}~\bibnamefont {Tamburino}},\ and\ \bibinfo {author} {\bibfnamefont {T.}~\bibnamefont {Unti}},\ }\bibfield  {title} {\bibinfo {title} {{Empty space generalization of the Schwarzschild metric}},\ }\href {https://doi.org/10.1063/1.1704018} {\bibfield  {journal} {\bibinfo  {journal} {J. Math. Phys.}\ }\textbf {\bibinfo {volume} {4}},\ \bibinfo {pages} {915} (\bibinfo {year} {1963})}\BibitemShut {NoStop}%
\bibitem [{\citenamefont {Harrison}(1968)}]{Harrison}%
  \BibitemOpen
  \bibfield  {author} {\bibinfo {author} {\bibfnamefont {B.}~\bibnamefont {Harrison}},\ }\bibfield  {title} {\bibinfo {title} {{New solutions of the Einstein-Maxwell equations from old}},\ }\href {https://doi.org/10.1063/1.1664508} {\bibfield  {journal} {\bibinfo  {journal} {J.Math.Phys.}\ }\textbf {\bibinfo {volume} {9}},\ \bibinfo {pages} {1744} (\bibinfo {year} {1968})}\BibitemShut {NoStop}%
\bibitem [{\citenamefont {Garfinkle}\ and\ \citenamefont {Melvin}(1994)}]{Garfinkle:1994wk}%
  \BibitemOpen
  \bibfield  {author} {\bibinfo {author} {\bibfnamefont {D.}~\bibnamefont {Garfinkle}}\ and\ \bibinfo {author} {\bibfnamefont {M.~A.}\ \bibnamefont {Melvin}},\ }\bibfield  {title} {\bibinfo {title} {{Generating new magnetic universe solutions from old}},\ }\href {https://doi.org/10.1103/PhysRevD.50.3859} {\bibfield  {journal} {\bibinfo  {journal} {Phys. Rev. D}\ }\textbf {\bibinfo {volume} {50}},\ \bibinfo {pages} {3859} (\bibinfo {year} {1994})},\ \Eprint {https://arxiv.org/abs/gr-qc/9404016} {arXiv:gr-qc/9404016} \BibitemShut {NoStop}%
\bibitem [{\citenamefont {Contopoulos}\ \emph {et~al.}(2015)\citenamefont {Contopoulos}, \citenamefont {Esposito}, \citenamefont {Kleidis}, \citenamefont {Papadopoulos},\ and\ \citenamefont {Witten}}]{Contopoulos:2015wra}%
  \BibitemOpen
  \bibfield  {author} {\bibinfo {author} {\bibfnamefont {I.~G.}\ \bibnamefont {Contopoulos}}, \bibinfo {author} {\bibfnamefont {F.~P.}\ \bibnamefont {Esposito}}, \bibinfo {author} {\bibfnamefont {K.}~\bibnamefont {Kleidis}}, \bibinfo {author} {\bibfnamefont {D.~B.}\ \bibnamefont {Papadopoulos}},\ and\ \bibinfo {author} {\bibfnamefont {L.}~\bibnamefont {Witten}},\ }\bibfield  {title} {\bibinfo {title} {{Generating Solutions to the Einstein Field Equations}},\ }\href {https://doi.org/10.1142/S021827181650022X} {\bibfield  {journal} {\bibinfo  {journal} {Int. J. Mod. Phys. D}\ }\textbf {\bibinfo {volume} {25}},\ \bibinfo {pages} {1650022} (\bibinfo {year} {2015})},\ \Eprint {https://arxiv.org/abs/1501.03968} {arXiv:1501.03968 [gr-qc]} \BibitemShut {NoStop}%
\bibitem [{\citenamefont {Astorino}\ \emph {et~al.}(2022)\citenamefont {Astorino}, \citenamefont {Martelli},\ and\ \citenamefont {Vigan{\`o}}}]{Astorino:2022aam}%
  \BibitemOpen
  \bibfield  {author} {\bibinfo {author} {\bibfnamefont {M.}~\bibnamefont {Astorino}}, \bibinfo {author} {\bibfnamefont {R.}~\bibnamefont {Martelli}},\ and\ \bibinfo {author} {\bibfnamefont {A.}~\bibnamefont {Vigan{\`o}}},\ }\bibfield  {title} {\bibinfo {title} {{Black holes in a swirling universe}},\ }\href {https://doi.org/10.1103/PhysRevD.106.064014} {\bibfield  {journal} {\bibinfo  {journal} {Phys. Rev. D}\ }\textbf {\bibinfo {volume} {106}},\ \bibinfo {pages} {064014} (\bibinfo {year} {2022})},\ \Eprint {https://arxiv.org/abs/2205.13548} {arXiv:2205.13548 [gr-qc]} \BibitemShut {NoStop}%
\bibitem [{\citenamefont {Barrientos}\ \emph {et~al.}(2024)\citenamefont {Barrientos}, \citenamefont {Cisterna}, \citenamefont {Kol{\'a}{\v{r}}}, \citenamefont {M{\"u}ller}, \citenamefont {Oyarzo},\ and\ \citenamefont {Pallikaris}}]{Barrientos:2024pkt}%
  \BibitemOpen
  \bibfield  {author} {\bibinfo {author} {\bibfnamefont {J.}~\bibnamefont {Barrientos}}, \bibinfo {author} {\bibfnamefont {A.}~\bibnamefont {Cisterna}}, \bibinfo {author} {\bibfnamefont {I.}~\bibnamefont {Kol{\'a}{\v{r}}}}, \bibinfo {author} {\bibfnamefont {K.}~\bibnamefont {M{\"u}ller}}, \bibinfo {author} {\bibfnamefont {M.}~\bibnamefont {Oyarzo}},\ and\ \bibinfo {author} {\bibfnamefont {K.}~\bibnamefont {Pallikaris}},\ }\bibfield  {title} {\bibinfo {title} {{Mixing {\textquotedblleft}Magnetic{\textquotedblright} and {\textquotedblleft}Electric{\textquotedblright} Ehlers{\textendash}Harrison transformations: the electromagnetic swirling spacetime and novel type I backgrounds}},\ }\href {https://doi.org/10.1140/epjc/s10052-024-13093-x} {\bibfield  {journal} {\bibinfo  {journal} {Eur. Phys. J. C}\ }\textbf {\bibinfo {volume} {84}},\ \bibinfo {pages} {724} (\bibinfo {year} {2024})},\ \Eprint {https://arxiv.org/abs/2401.02924} {arXiv:2401.02924 [gr-qc]} \BibitemShut {NoStop}%
\bibitem [{\citenamefont {Di~Pinto}\ \emph {et~al.}(2025)\citenamefont {Di~Pinto}, \citenamefont {Klemm},\ and\ \citenamefont {Vigan{\`o}}}]{DiPinto:2025yaa}%
  \BibitemOpen
  \bibfield  {author} {\bibinfo {author} {\bibfnamefont {A.}~\bibnamefont {Di~Pinto}}, \bibinfo {author} {\bibfnamefont {S.}~\bibnamefont {Klemm}},\ and\ \bibinfo {author} {\bibfnamefont {A.}~\bibnamefont {Vigan{\`o}}},\ }\bibfield  {title} {\bibinfo {title} {{Kerr-Newman black hole in a Melvin-swirling universe}},\ }\href {https://doi.org/10.1007/JHEP06(2025)150} {\bibfield  {journal} {\bibinfo  {journal} {JHEP}\ }\textbf {\bibinfo {volume} {06}},\ \bibinfo {pages} {150}},\ \Eprint {https://arxiv.org/abs/2503.07780} {arXiv:2503.07780 [gr-qc]} \BibitemShut {NoStop}%
\bibitem [{\citenamefont {Cisterna}\ \emph {et~al.}(2023)\citenamefont {Cisterna}, \citenamefont {M{\"u}ller}, \citenamefont {Pallikaris},\ and\ \citenamefont {Vigan{\`o}}}]{Cisterna:2023uqf}%
  \BibitemOpen
  \bibfield  {author} {\bibinfo {author} {\bibfnamefont {A.}~\bibnamefont {Cisterna}}, \bibinfo {author} {\bibfnamefont {K.}~\bibnamefont {M{\"u}ller}}, \bibinfo {author} {\bibfnamefont {K.}~\bibnamefont {Pallikaris}},\ and\ \bibinfo {author} {\bibfnamefont {A.}~\bibnamefont {Vigan{\`o}}},\ }\bibfield  {title} {\bibinfo {title} {{Exact rotating wormholes via Ehlers transformations}},\ }\href {https://doi.org/10.1103/PhysRevD.108.024066} {\bibfield  {journal} {\bibinfo  {journal} {Phys. Rev. D}\ }\textbf {\bibinfo {volume} {108}},\ \bibinfo {pages} {024066} (\bibinfo {year} {2023})},\ \Eprint {https://arxiv.org/abs/2306.14541} {arXiv:2306.14541 [gr-qc]} \BibitemShut {NoStop}%
\bibitem [{\citenamefont {Capobianco}\ \emph {et~al.}(2024{\natexlab{a}})\citenamefont {Capobianco}, \citenamefont {Hartmann},\ and\ \citenamefont {Kunz}}]{Capobianco:2023kse}%
  \BibitemOpen
  \bibfield  {author} {\bibinfo {author} {\bibfnamefont {R.}~\bibnamefont {Capobianco}}, \bibinfo {author} {\bibfnamefont {B.}~\bibnamefont {Hartmann}},\ and\ \bibinfo {author} {\bibfnamefont {J.}~\bibnamefont {Kunz}},\ }\bibfield  {title} {\bibinfo {title} {{Geodesic motion in a swirling universe: The complete set of solutions}},\ }\href {https://doi.org/10.1103/PhysRevD.109.064042} {\bibfield  {journal} {\bibinfo  {journal} {Phys. Rev. D}\ }\textbf {\bibinfo {volume} {109}},\ \bibinfo {pages} {064042} (\bibinfo {year} {2024}{\natexlab{a}})},\ \Eprint {https://arxiv.org/abs/2312.17347} {arXiv:2312.17347 [gr-qc]} \BibitemShut {NoStop}%
\bibitem [{\citenamefont {Capobianco}\ \emph {et~al.}(2024{\natexlab{b}})\citenamefont {Capobianco}, \citenamefont {Hartmann},\ and\ \citenamefont {Kunz}}]{Capobianco:2024jhe}%
  \BibitemOpen
  \bibfield  {author} {\bibinfo {author} {\bibfnamefont {R.}~\bibnamefont {Capobianco}}, \bibinfo {author} {\bibfnamefont {B.}~\bibnamefont {Hartmann}},\ and\ \bibinfo {author} {\bibfnamefont {J.}~\bibnamefont {Kunz}},\ }\bibfield  {title} {\bibinfo {title} {{Motion of charged particles in an electromagnetic swirling universe: The complete set of solutions}},\ }\href {https://doi.org/10.1103/PhysRevD.110.084078} {\bibfield  {journal} {\bibinfo  {journal} {Phys. Rev. D}\ }\textbf {\bibinfo {volume} {110}},\ \bibinfo {pages} {084078} (\bibinfo {year} {2024}{\natexlab{b}})},\ \Eprint {https://arxiv.org/abs/2407.03702} {arXiv:2407.03702 [gr-qc]} \BibitemShut {NoStop}%
\bibitem [{\citenamefont {Moreira}\ \emph {et~al.}(2024)\citenamefont {Moreira}, \citenamefont {Herdeiro},\ and\ \citenamefont {Crispino}}]{Moreira:2024sjq}%
  \BibitemOpen
  \bibfield  {author} {\bibinfo {author} {\bibfnamefont {Z.~S.}\ \bibnamefont {Moreira}}, \bibinfo {author} {\bibfnamefont {C.~A.~R.}\ \bibnamefont {Herdeiro}},\ and\ \bibinfo {author} {\bibfnamefont {L.~C.~B.}\ \bibnamefont {Crispino}},\ }\bibfield  {title} {\bibinfo {title} {{Twisting shadows: Light rings, lensing, and shadows of black holes in swirling universes}},\ }\href {https://doi.org/10.1103/PhysRevD.109.104020} {\bibfield  {journal} {\bibinfo  {journal} {Phys. Rev. D}\ }\textbf {\bibinfo {volume} {109}},\ \bibinfo {pages} {104020} (\bibinfo {year} {2024})},\ \bibinfo {note} {[Erratum: Phys.Rev.D 112, 109903 (2025)]},\ \Eprint {https://arxiv.org/abs/2401.05658} {arXiv:2401.05658 [gr-qc]} \BibitemShut {NoStop}%
\bibitem [{\citenamefont {Capobianco}\ \emph {et~al.}(2026)\citenamefont {Capobianco}, \citenamefont {Hartmann}, \citenamefont {Vas}, \citenamefont {Kunz},\ and\ \citenamefont {Novo}}]{Capobianco:2025lxn}%
  \BibitemOpen
  \bibfield  {author} {\bibinfo {author} {\bibfnamefont {R.}~\bibnamefont {Capobianco}}, \bibinfo {author} {\bibfnamefont {B.}~\bibnamefont {Hartmann}}, \bibinfo {author} {\bibfnamefont {N.}~\bibnamefont {Vas}}, \bibinfo {author} {\bibfnamefont {J.}~\bibnamefont {Kunz}},\ and\ \bibinfo {author} {\bibfnamefont {J.}~\bibnamefont {Novo}},\ }\bibfield  {title} {\bibinfo {title} {{Photon rings and shadows of Kerr black holes immersed in a swirling universe}},\ }\href {https://doi.org/10.1103/wj2m-s3sv} {\bibfield  {journal} {\bibinfo  {journal} {Phys. Rev. D}\ }\textbf {\bibinfo {volume} {113}},\ \bibinfo {pages} {064053} (\bibinfo {year} {2026})},\ \Eprint {https://arxiv.org/abs/2510.01937} {arXiv:2510.01937 [gr-qc]} \BibitemShut {NoStop}%
\bibitem [{\citenamefont {Barrientos}\ \emph {et~al.}(2025)\citenamefont {Barrientos}, \citenamefont {Cisterna}, \citenamefont {Hassaine},\ and\ \citenamefont {Pallikaris}}]{Barrientos:2024umq}%
  \BibitemOpen
  \bibfield  {author} {\bibinfo {author} {\bibfnamefont {J.}~\bibnamefont {Barrientos}}, \bibinfo {author} {\bibfnamefont {A.}~\bibnamefont {Cisterna}}, \bibinfo {author} {\bibfnamefont {M.}~\bibnamefont {Hassaine}},\ and\ \bibinfo {author} {\bibfnamefont {K.}~\bibnamefont {Pallikaris}},\ }\bibfield  {title} {\bibinfo {title} {{Electromagnetized black holes and swirling backgrounds in nonlinear electrodynamics: The ModMax case}},\ }\href {https://doi.org/10.1016/j.physletb.2024.139214} {\bibfield  {journal} {\bibinfo  {journal} {Phys. Lett. B}\ }\textbf {\bibinfo {volume} {860}},\ \bibinfo {pages} {139214} (\bibinfo {year} {2025})},\ \Eprint {https://arxiv.org/abs/2409.12336} {arXiv:2409.12336 [gr-qc]} \BibitemShut {NoStop}%
\bibitem [{\citenamefont {Coll{\'e}aux}\ \emph {et~al.}(2026)\citenamefont {Coll{\'e}aux}, \citenamefont {Kol{\'a}{\v{r}}},\ and\ \citenamefont {M{\'a}lek}}]{Colleaux:2026qew}%
  \BibitemOpen
  \bibfield  {author} {\bibinfo {author} {\bibfnamefont {A.}~\bibnamefont {Coll{\'e}aux}}, \bibinfo {author} {\bibfnamefont {I.}~\bibnamefont {Kol{\'a}{\v{r}}}},\ and\ \bibinfo {author} {\bibfnamefont {T.}~\bibnamefont {M{\'a}lek}},\ }\href@noop {} {\bibinfo {title} {{Quasi-topological gravity for 4-dimensional Taub-NUT, near-horizon extreme Kerr, and swirling symmetries}}} (\bibinfo {year} {2026}),\ \Eprint {https://arxiv.org/abs/2606.17784} {arXiv:2606.17784 [gr-qc]} \BibitemShut {NoStop}%
\bibitem [{\citenamefont {Giusto}\ and\ \citenamefont {Saxena}(2007)}]{Giusto:2007fx}%
  \BibitemOpen
  \bibfield  {author} {\bibinfo {author} {\bibfnamefont {S.}~\bibnamefont {Giusto}}\ and\ \bibinfo {author} {\bibfnamefont {A.}~\bibnamefont {Saxena}},\ }\bibfield  {title} {\bibinfo {title} {{Stationary axisymmetric solutions of five dimensional gravity}},\ }\href {https://doi.org/10.1088/0264-9381/24/17/002} {\bibfield  {journal} {\bibinfo  {journal} {Class. Quant. Grav.}\ }\textbf {\bibinfo {volume} {24}},\ \bibinfo {pages} {4269} (\bibinfo {year} {2007})},\ \Eprint {https://arxiv.org/abs/0705.4484} {arXiv:0705.4484 [hep-th]} \BibitemShut {NoStop}%
\bibitem [{\citenamefont {Suzuki}\ and\ \citenamefont {Tomizawa}(2024)}]{Suzuki:2024eoz}%
  \BibitemOpen
  \bibfield  {author} {\bibinfo {author} {\bibfnamefont {R.}~\bibnamefont {Suzuki}}\ and\ \bibinfo {author} {\bibfnamefont {S.}~\bibnamefont {Tomizawa}},\ }\bibfield  {title} {\bibinfo {title} {{New construction of a vacuum doubly rotating black ring by the Ehlers transformation}},\ }\href {https://doi.org/10.1103/PhysRevD.110.084024} {\bibfield  {journal} {\bibinfo  {journal} {Phys. Rev. D}\ }\textbf {\bibinfo {volume} {110}},\ \bibinfo {pages} {084024} (\bibinfo {year} {2024})},\ \Eprint {https://arxiv.org/abs/2406.11804} {arXiv:2406.11804 [hep-th]} \BibitemShut {NoStop}%
\bibitem [{\citenamefont {Yazadjiev}(2008)}]{Yazadjiev:2008pt}%
  \BibitemOpen
  \bibfield  {author} {\bibinfo {author} {\bibfnamefont {S.~S.}\ \bibnamefont {Yazadjiev}},\ }\bibfield  {title} {\bibinfo {title} {{5D Einstein-Maxwell solitons and concentric rotating dipole black rings}},\ }\href {https://doi.org/10.1103/PhysRevD.78.064032} {\bibfield  {journal} {\bibinfo  {journal} {Phys. Rev. D}\ }\textbf {\bibinfo {volume} {78}},\ \bibinfo {pages} {064032} (\bibinfo {year} {2008})},\ \Eprint {https://arxiv.org/abs/0805.1600} {arXiv:0805.1600 [hep-th]} \BibitemShut {NoStop}%
\bibitem [{\citenamefont {Vigan{\`o}}(2026)}]{Vigano:2026hqd}%
  \BibitemOpen
  \bibfield  {author} {\bibinfo {author} {\bibfnamefont {A.}~\bibnamefont {Vigan{\`o}}},\ }\href@noop {} {\bibinfo {title} {{Black objects in a five-dimensional swirling spacetime}}} (\bibinfo {year} {2026}),\ \Eprint {https://arxiv.org/abs/2609.37511} {arXiv:2609.37511 [gr-qc]} \BibitemShut {NoStop}%
\bibitem [{\citenamefont {Barrientos~et al.}(2026)}]{toappear}%
  \BibitemOpen
  \bibfield  {author} {\bibinfo {author} {\bibfnamefont {J.}~\bibnamefont {Barrientos~et al.}},\ }\bibfield  {title} {\bibinfo {title} {{General 5D Swirling Myers-Perry geometries}},\ }\href@noop {} {\  (\bibinfo {year} {2026})}\BibitemShut {NoStop}%
\bibitem [{\citenamefont {{Hawking}}(1969)}]{Hawking1969}%
  \BibitemOpen
  \bibfield  {author} {\bibinfo {author} {\bibfnamefont {S.~W.}\ \bibnamefont {{Hawking}}},\ }\bibfield  {title} {\bibinfo {title} {{The Existence of Cosmic Time Functions}},\ }\href {https://doi.org/10.1098/rspa.1969.0018} {\bibfield  {journal} {\bibinfo  {journal} {Proceedings of the Royal Society of London Series A}\ }\textbf {\bibinfo {volume} {308}},\ \bibinfo {pages} {433} (\bibinfo {year} {1969})}\BibitemShut {NoStop}%
\bibitem [{\citenamefont {Abramowitz}\ and\ \citenamefont {Stegun}(1964)}]{abramowitz1964handbook}%
  \BibitemOpen
  \bibfield  {author} {\bibinfo {author} {\bibfnamefont {M.}~\bibnamefont {Abramowitz}}\ and\ \bibinfo {author} {\bibfnamefont {I.~A.}\ \bibnamefont {Stegun}},\ }\href@noop {} {\emph {\bibinfo {title} {Handbook of Mathematical Functions with Formulas, Graphs, and Mathematical Tables}}},\ Vol.~\bibinfo {volume} {55}\ (\bibinfo  {publisher} {National Bureau of Standards / Dover Publications},\ \bibinfo {year} {1964})\BibitemShut {NoStop}%
\end{thebibliography}%
\end{document}